\documentclass[acmsmall]{acmart}

\usepackage{url}
\AtBeginDocument{%
  \providecommand\BibTeX{{%
    \normalfont B\kern-0.5em{\scshape i\kern-0.25em b}\kern-0.8em\TeX}}}

\begin{document}

\title[Relevance Assessments for Web Search Evaluation (CORRECTED)]{Relevance Assessments for Web Search Evaluation:
Should We Randomise or Prioritise the Pooled Documents? (CORRECTED VERSION)}

\author{Tetsuya Sakai}
\email{tetsuyasakai@acm.org}
\author{Sijie Tao}
\email{tsjmailbox@ruri.waseda.jp}
\author{Zhaohao Zeng}
\email{zhaohao@fuji.waseda.jp}
\authornotemark[1]
%\email{webmaster@marysville-ohio.com}
\affiliation{%
  \institution{Waseda University}
  \streetaddress{3-4-1 Okubo, Shinjuku}
  \city{Tokyo}
  \country{Japan}
  \postcode{169-8555}
}

%\author{Lars Th{\o}rv{\"a}ld}
%\affiliation{%
%  \institution{The Th{\o}rv{\"a}ld Group}
%  \streetaddress{1 Th{\o}rv{\"a}ld Circle}
%\email{larst@affiliation.org}

%}


\begin{abstract}
In the context of depth-$k$ pooling
for constructing web search test collections,
we compare two approaches to
ordering pooled documents for relevance assessors:
the prioritisation strategy (PRI) used widely at NTCIR,
and the simple randomisation strategy (RND).
In order to address research questions regarding PRI and RND,
we have constructed and released the 
WWW3E8 data set,
which contains eight independent relevance labels 
for 32,375 topic-document pairs,
i.e., a total of 259,000 labels.
Four of the eight relevance labels were obtained 
from PRI-based pools;
the other four were obtained 
from RND-based pools.
Using WWW3E8,
we compare PRI and RND in terms of inter-assessor agreement,
system ranking agreement, 
and robustness to new systems that did not contribute to the pools.
We also utilise an assessor activity log we obtained as a byproduct of WWW3E8
to compare the two strategies in terms of assessment efficiency.
Our main findings are as follows.
%(a)~the presentation order
%has no substantial impact on assessment efficiency;
(a)~There is no substantial difference between RND and PRI
in terms of time spent for judging each document,
although PRI may enable
faster identification of the first highly relevant document in the pool.
%(b)~while the presentation order substantially affects
%which documents are judged (highly) relevant, 
%the difference between the inter-assessor agreement under the PRI condition
%and that under the RND condition
% is of no practical significance;
(b)~The difference between
the inter-assessor agreement under the RND condition
and that under the PRI condition is probably of no practical significance.
%(c)~different system rankings under the PRI condition are 
%substantially more similar to one another
%than those under the RND condition; and
(c)~While PRI-based qrels files tend to generate system ranking that are slightly more similar to each other
than RND-based qrels files do,
this difference is probably of no practical significance.
On the other hand, 
a PRI-based system ranking and a RND-based system ranking
can be quite different relative to the above within-PRI and within-RND ranking comparisons.
The PRI strategy tends to make the assessor favour ``popular'' documents,
i.e., those returned at high ranks by many systems.
%(d)~PRI-based relevance assessment files (qrels)
%are substantially and statistically significantly more robust
%to new systems than RND-based ones.
%Finding~(d) suggests that
%PRI helps the assessors identify 
%relevant documents that affect the evaluation of many existing systems,
%including those that did not contribute to the pools.
%Hence, 
%if researchers need to evaluate their current IR systems using legacy IR test collections,
%we recommend the use of those constructed using the PRI approach
%unless they have a good reason to believe that
%their systems retrieve relevant documents that are vastly different
%from the pooled documents.
%While this robustness of PRI may also mean 
%that the PRI-based pools are biased against future systems
%that retrieve highly novel relevant documents,
%one should note that
%there is no evidence that RND is any better in this respect.
(d)~PRI-based qrels files tend to be slightly more robust to new systems than RND-based ones.
This is probably because the PRI strategy tends to help us identify ``popular'' relevant documents.
The ``popular'' relevant documents affect the evaluation of many systems,
including systems that did not contribute to the pools.
That is, our results suggests that PRI-based test collections
may be slightly more reusable than RND-based ones.\footnote{
This paper is a corrected version of our open-access TOIS paper which suffered from a major bug (as well as a minor one) as explained in the
corrigendum~\cite{sakai22tois}. 
The nature of the major bug is also explained in the corrected NTCIR We Want Web results paper~\cite{sakai22www234corrected}.
We apologise to the TOIS reviewers, editors, and those who have already read our TOIS paper,
for the confusion and inconveniences that we have inadvertently caused.
We thank the editor-in-chief for allowing us to publish this revised version on arxiv in addition to the official
corrigendum to the TOIS paper.
}
\end{abstract}

%\begin{CCSXML}
%<ccs2012>
%<concept>
%<concept_id>10002951.10003317.10003359.10003360</concept_id>
%<concept_desc>Information systems~Test collections</concept_desc>
%<concept_significance>500</concept_significance>
%</concept>
%<concept>
%<concept_id>10002951.10003317.10003359.10003361</concept_id>
%<concept_desc>Information systems~Relevance assessment</concept_desc>
%<concept_significance>500</concept_significance>
%</concept>
%<concept>
%<concept_id>10002951.10003317.10003359.10003362</concept_id>
%<concept_desc>Information systems~Retrieval effectiveness</concept_desc>
%<concept_significance>500</concept_significance>
%</concept>
%</ccs2012>
%\end{CCSXML}

%\ccsdesc[500]{Information systems~Test collections}
%\ccsdesc[500]{Information systems~Relevance assessment}
%\ccsdesc[500]{Information systems~Retrieval effectiveness}


%\setcopyright{acmlicensed}
%\acmJournal{TOIS}
% edit this
%\acmYear{2021} \acmVolume{1} \acmNumber{1} \acmArticle{1} \acmMonth{1} \acmPrice{15.00}\acmDOI{10.1145/3494833}

%\received{May 2021}
%\received[revised]{August 2021}
%\received[accepted]{October 2021}

\maketitle

\section{Introduction}\label{s:intro}

Decades after the Cranfield II experiments based on relevance assessments in the 1960s~\cite{cleverdon66a,cleverdon66b}
and the proposal of pooling-based `ideal' test collections in the 1970s~\cite{sparckjones75,sparckjones77},
offline information retrieval system evaluation
using pooling-based test collections still remains
vital for
providing researchers with insight into why some retrieval methods work while others do not,
and for helping them advance the state-of-the-art by building on top of that knowledge.
More specifically, 
\emph{depth-$k$} pooling,
the approach 
of taking top $k$ documents from each participating run to form a pool
for relevance assessments~\cite{harman05,sakai19clefbook},
has played a central role in collecting relevance assessments for test collections with large document collections,
ever 
since the advent of TREC\footnote{Text Retrieval Conference~\cite{harman05}.} in the early 1990s.
The present study concerns how the pooled documents are ordered and presented to the 
relevance assessors; 
in the context of depth-$k$ pooling,
we 
address a few research questions that have been debated over the past two decades or so, as discussed more specifically below.

While many IR tasks still rely on depth-$k$ pooling, 
the pooled documents are ordered for the assessors differently at different tasks.
In a paper given at CLEF\footnote{
Cross-Language Evaluation Forum, now known as 
Conference and Labs of the Evaluation Forum~\cite{ferro19book},
} 2001,
Voorhees explains the early TREC approach~\cite[p.357]{voorhees02}:
``{\it Each pool is sorted by document identifier
so assessors cannot tell if a document was
highly ranked by some system or how many systems (or which systems)
retrieved the document.}''
An almost identical comment by Harman can be found in the TREC book published in 2005~\cite[p.33]{harman05}.
The philosophy there is
to \emph{randomise} the document judging order
to avoid judgement biases:
for example, if the assessor knows that
a document was retrieved by many participating systems
at top ranks, then this knowledge might 
make them overestimate its relevance.
We call this the randomisation approach (or \emph{RND} for brevity).
Note that we do not take of the above advice of
``sorting by document number'' literally:
if the publication date is embedded in the document identifier,
then sorting by document ID would mean sorting by time, which is 
not what we want. 
Similarly, 
as \citet{damessie18} have observed,
if the target document collection consists of 
multiple subcollections and the document IDs contain
different prefixes accordingly,
such a sort would actually cluster documents by source,
which again is not what we want.

Interestingly, many pooling-based IR tasks of NTCIR\footnote{
NII (National Institute of Informatics) Testbeds and Community for Information access Research~\cite{sakai20book}.
}
do not follow the above advice from TREC. 
In fact, they do the exact opposite.
At CLEF 2003,
Kando explained~\cite[p.38]{kando04}:
``{\it Pooled document lists to be judged are sorted in descending order of likelihood of being relevant (not the order of the document IDs)}''
and remarks that the NTCIR's decision was 
``{\it based on comparative tests and interviews with assessors}.''
However, the details were never published.
In 2008, \citet{sakai08ir4qa}
introduced a specific implementation of NTCIR's ordering approach,
now widely used at NTCIR with the {\tt NTCIRPOOL} tool~\cite{sakai19clefbook}\footnote{
\url{http://research.nii.ac.jp/ntcir/tools/ntcirpool-en.html}
}:
the pooled documents are sorted by \emph{pseudorelevance},
where
the first sort key is
the number of runs containing the document
at or above the pool depth $k$ (the larger the better),
and
the second sort key is
the sum of ranks of that document within those runs
(the smaller the better).
Thus, documents that were retrieved by many systems 
at high ranks are \emph{prioritised}.
This NTCIR approach, hereafter referred to as \emph{PRI},
aims to let the assessors 
go through the documents roughly in decreasing order of relevance
so that they can quickly form an idea as to what constitutes a relevant document, 
and thereby enhance \emph{assessment efficiency} and \emph{inter-assessor agreement}~\cite{sakai19airs}.
To date, however, there is no concrete evidence that
supports the above claims.

RND and PRI are probably the two most simple
and widely-used document ordering strategies for constructing 
depth-$k$ pool-based IR test collections.
Even after the past two decades or so,\footnote{
The first NTCIR conference (workshop) was held in 1999~\cite{sakai20book}.
} however,
the IR research community has yet to reach a consensus  as to
what advantages each of these two strategies actually offer.
Hence, the present study addresses the following research questions for PRI and RND.
\begin{description}
\item[RQ1] Which strategy enables more efficient relevance assessments?
\item[RQ2] Which strategy enables higher inter-assessor agreements?
\item[RQ3] Which strategy enables more stable system rankings across different versions of qrels files?
\item[RQ4] Which strategy is more robust to the evaluation of systems that did not contribute to the pools?
\end{description}
In order to address research questions regarding PRI and RND,
we have constructed and released the 
WWW3E8 data set~\cite{sakai21www3e8},\footnote{
Available at \url{https://waseda.box.com/WWW3E8corrected}
}
which contains eight independent relevance labels 
for 32,375 topic-document pairs,
i.e., a total of 259,000 labels.
Four of the eight relevance labels were obtained 
from PRI-based pools;
the other four were obtained 
from RND-based pools.
Using WWW3E8,
we compare PRI and RND in terms of inter-assessor agreement,
system ranking agreement, 
and robustness to new systems that did not contribute to the pools.
We also utilise an assessor activity log we obtained as a byproduct of WWW3E8
to compare the two strategies in terms of assessment efficiency.
Our main findings are as follows.
%(a)~the presentation order
%has no substantial impact on assessment efficiency;
%(b)~while the presentation order substantially affects
%which documents are judged (highly) relevant, 
%the difference between the inter-assessor agreement under the PRI condition
%and that under the RND condition
% is of no practical significance;
%(c)~different system rankings under the PRI condition are 
%substantially more similar to one another
%than those under the RND condition; and
%(d)~PRI-based relevance assessment files (qrels)
%are substantially and statistically significantly more robust
%to new systems than RND-based ones.
%Finding~(d) suggests that
%PRI helps the assessors identify 
%relevant documents that affect the evaluation of many systems,
%including those that did not contribute to the pools.
%Hence, 
%if researchers need to evaluate their current IR systems using legacy IR test collections,
%we recommend the use of those constructed using the PRI approach
%unless they have a good reason to believe that
%their systems retrieve relevant documents that are vastly different
%from the pooled documents.
%Put another way,
%if the researchers believe that
%their new systems return search results 
%that are reasonably similar to existing systems,
%then a PRI-based test collection is recommended;
%otherwise the new systems may be heavily underrated.
%While this robustness of PRI may also mean 
%that the PRI-based pools are biased against future systems
%that retrieve highly novel relevant documents,
%one should note that
%there is no evidence that RND is any better in this respect.
(a)~There is no substantial difference between RND and PRI
in terms of time spent for judging each document,
although PRI may enable
faster identification of the first highly relevant document in the pool.
(b)~The difference between
the inter-assessor agreement under the RND condition
and that under the PRI condition is probably of no practical significance.
(c)~While PRI-based qrels files tend to generate system ranking that are slightly more similar to each other
than RND-based qrels files do,
this difference is probably of no practical significance.
On the other hand, 
a PRI-based system ranking and a RND-based system ranking
can be quite different relative to the above within-PRI and within-RND ranking comparisons.
The PRI strategy tends to make the assessor favour ``popular'' documents,
i.e., those returned at high ranks by many systems.
(d)~PRI-based qrels files tend to be slightly more robust to new systems than RND-based ones.
This is probably because the PRI strategy tends to help us identify ``popular'' relevant documents.
The ``popular'' relevant documents affect the evaluation of many systems,
including systems that did not contribute to the pools.
That is, our results suggests that PRI-based test collections
may be slightly more reusable than RND-based ones.

%The present study extends our SIGIR 2021 resource paper that introduced the WWW3E8 data set~\cite{sakai21www3e8},
%by reporting on extensive experiments conducted by utilising WWW3E8.
%We acknowledge that Sections~\ref{s:related}-\ref{s:WWW3E8} basically
%duplicate the content of our resource paper:
%this is to ensure that the present study is self-contained as a journal paper. 
%The major component of the present paper is the examination of the four research questions mentioned above.
%We also acknowledge that the relevance assessments of the WWW3E8 data set
%were obtained from \emph{bronze} assessors~\cite{bailey08},
%i.e., non-experts who are not the creators of the search topics (See Section~\ref{s:related});
%we shall discuss how we plan to follow up on this limitation in Section~\ref{s:conclusions}.

%The present study is also a follow-up study of the work by Sakai and Xiao~\cite{sakai19airs},
%who tried to address \textbf{RQ1} and \textbf{RQ2} but reported 
%that the statistical powers of their experiments were too low.
%The design of our WWW3E8 data set was directly based on the sample size considerations
%discussed by Sakai and Xiao.
%We shall discuss the differences between their work and ours in Section~\ref{ss:airs2019}.

\section{Prior Art}\label{s:related}

Test collection-based evaluations of IR systems depend on human relevance assessments
and therefore ensuring the reliability of the assessments as the ground truth
is of utmost importance to the IR community.
Accordingly, there is a large body of work on the reliability of relevance assessments.
For example, \citet{voorhees00} demonstrated 
that, while different assessors rate the same documents somewhat differently,
when we rank systems by mean effectiveness scores,
the system rankings are quite robust to the change in the set of relevance assessments.
\citet{bailey08} examined the effect of assessor expertise
on ranking systems, by considering three types of relevance assessors:
\emph{gold} (topic originators), \emph{silver} (task experts who are not topic originators),
and \emph{bronze} (who are neither). 
From this viewpoint,
Cranfield II~\cite{cleverdon66a,cleverdon66b}, the first IR experiment that
involved relevance assessments, involved gold assessors (in addition to students);
the present study concerns bronze assessors, as we shall discuss in Section~\ref{s:reliability}.
Below, we focus our attention on prior art on document selection strategies for pooling (Section~\ref{ss:alternatives}),
and those on document ordering for relevance assessors (Section~\ref{ss:ordering}).
%Regarding document ordering, Section~\ref{ss:airs2019} specifically
%discusses the work of Sakai and Xiao~\cite{sakai19airs},
%as their sample size considerations were utilised 
%for the design of our WWW3E8 data set;
%we also clarify the differences between their work and ours.

\subsection{Alternatives to Depth-$k$ Pooling}\label{ss:alternatives}

Although several alternatives to depth-$k$ pooling have been proposed,
it is probably fair to say that none are as widely-used.
In 1998,
\citet{zobel98} proposed to allocate more judging resources
to more promising topics, while
\citet{cormack98} proposed to allocate more judging resources 
to more promising runs.
TREC did not adopt these methods for
fear of introducing bias~\cite{voorhees02}.
The TREC Million Query Track~\cite{allan08,carterette10trec} used
the \emph{Minimal Test Collections} (MTC) and \emph{statAP} methods.
MTC iteratively orders documents to be judged
according to how much information they provide about 
a difference in average precision;
statAP samples documents to be judged
based on a sampling distribution 
that tries to place higher weights on relevant documents.
More recently, 
the TREC 2017 Common Core Track~\cite{allan18}
re-examined the problem of 
how best to go beyond depth-$k$ pooling.
They adopted a version of 
the \emph{MaxMean} method of \citet{losada17},
which dynamically selects which run to process
based on the judgements so far;
this is similar in spirit to 
the aforementioned method of Cormack {\it et al.}
in that it tries to focus
judging resources on 
those runs that continue to contribute relevant documents.
The Common Core Track 2017 overview paper
mentions \emph{judgement bias}, i.e.,
the bias caused by presenting top-ranked documents to the assessors first,
and \emph{run bias}, i.e.,
underestimating runs that do not contribute many relevant documents 
early in their rankings.
While
the overview paper reports
that there was no indication of run bias in their experiments,
judgement bias was not tested.

We are primarily interested in depth-$k$ pooling 
because our view is that, generally speaking, 
dynamic document selection approaches
have a few practical inconveniences when compared to depth-$k$ pooling.
Firstly, they require that once a relevance assessment is made, that is final.
In contrast, with depth-$k$ pooling, assessors can correct their judgements at any time in any order, as long as the judgement interface allows; the present study actually examines how often assessors correct their labels.
Secondly, dynamic approaches complicate logistics:
for example, compared to depth-$k$ pooling,
it may be more difficult to anticipate the workload per assessor,
and to let multiple assessors handle the same topic.
In short, we view the static nature of depth-$k$ pooling as a strength.

In the aforementioned Common Core Track~\cite{allan18},
the track coordinators point out that
it is important to give each assessor
an initial \emph{burn-in period},
in which  the assessor learns about the topic and
optionally make changes to their own initial relevance assessments;
hence the coordinators implemented a hybrid approach
where the assessor initially processes a traditional depth-10 pool
and then moves to the MaxMean-based dynamic judging phase.
That is, even this TREC track actually relies on depth-$k$ pooling.
We also note that the aim of the burn-in period resembles NTCIR's motivation for prioritising the pooled documents
(See Section~\ref{s:intro}).

The recent work of \citet{lipani21}
provides a comprehensive study on alternatives to depth-$k$ pooling,
including adaptive (i.e., dynamic) methods such as the ones discussed above~\cite{losada17} as well as nonadaptive methods
adopted from ranking fusion,
i.e., computing pseudorelevance scores based on multiple ranked lists~\cite{sakai10evia}.
For nonadaptive settings,
Lipani \textit{et al.} recommend
using the maximum score obtained by a document across all runs
as an alternative to depth-$k$ pooling.
However, this is beyond the scope of the present study.

\subsection{Document Ordering for Assessors}\label{ss:ordering}

Document judging order was a concern even before TREC. 
For example, in 1988,
\citet{eisenberg88} reported on
a small-scale experiment
where 15 document descriptions for a single topic were presented (on paper) to each assessor 
in either increasing or decreasing order of relevance,
and the assessors were asked to rate the documents on a 7-point scale.
They observed a \emph{hedging phenomenon}:
the assessors were reluctant to label the early documents with 
very high or very low scores, because they might want to reserve these extreme scores
for later documents.\footnote{
\citet{eisenberg88} considered \emph{magnitude estimation} (See also \citet{turpin15})
as an alternative to 7-point ratings; this is beyond the scope of our study.
}
Thus, for example, when the documents were presented in decreasing order of relevance,
the assessors tended to underestimate the relevance of the early documents.
Based on their results, they caution against judgement bias
and recommend randomising the presentation order.
In 2004, \citet{huang04}
reported on a similar experiment with similar recommendations,
but varied the number of documents to judge: $d=5, 15, 30, 45, 60, 75$.
They reported that order effects were not observed when $d=5$ and $d=75$,
and conjectured that the latter result may be due to fatigue: ``\textit{an excess
of documents simply exhausts the subjects with the toil and leads to responding without careful
considerations.}''
%s 75 or greater. Note that, in modern large-scale depth-$k$ pooling-based evaluations,
%the number of documents to judge per topic is typically several \emph{hundreds} (See Section~\ref{s:constructing}).

We do not consider the above studies as directly comparable to our work 
for the following reasons.
\begin{itemize}
\item They used a 7-point scale ratings, which is probably the primary reason for the hedging phenomenon:
it is probably difficult to give a 7-point label or a 1-point label to an early document without having seen the other documents.
In contrast, our relevance assessment task is simpler, as the choices are \textbf{highly relevant}, \textbf{relevant}, or \textbf{nonrelevant} (and \textbf{error})
as described in Section~\ref{ss:constructing}.
\item Their relevance assessors examined the documents provided in paper form (even in the 2004 paper of \citet{huang04});
we use a web browser-based interface, which allows the assessors to go back and forth on the document list and even correct their relevance labels if they wanted to.
\item Their experimental designs are based on the implicit assumption that the RND-based relevance assessments are perfect and correct:
that is, they obtained the relevance assessments first based on the RND document ordering,
and then sorted the judged documents based on the RND-based relevance levels to examine the order effect.
In contrast, we make no such assumption: 
we directly compare the RND and PRI conditions and study which documents are judged relevant, as well as the outcomes of these differences.
\item Their experiments relied on a single search topic; we have 160 different topics.
\end{itemize}
Note also that while  \citet{huang04} discussed the possibility of fatigue with $d=75$ documents (in paper form),
modern relevance assessments (with documents presented on a computer screen) typically involves hundreds of documents per topic.
In our study, each assessor handled 53-54 topics, where each topic has about 202 documents on average. While fatigue may well be 
playing a part in our experiments, this is beyond the scope of our study,
as how each assessor works  (e.g., how many topics or documents they process each day)  during the two months that they were given
was completely at their discretion  (See Section~\ref{ss:constructing}).

In 2013, \citet{scholer13} studied the effect of the overall relevance of early documents
on the 4-point assessor ratings of later documents, using three topics from TREC
and 48 documents per topic.
The first 20 documents presented to the assessors were called the Prologue; the other 28 were called the Epilogue.
They controlled the quality of the Prologue based on existing relevance assessments (treated as the gold standard).\footnote{
Unlike the earlier studies of \citet{eisenberg88} and \citet{huang04},
the work of Scholer \textit{et al.} used a relevance assessment interface as in our study.}
By comparing across the three Prologues conditions (high/medium/low relevance),
they observed an effect similar to the hedging phenomenon of \citet{eisenberg88},
and argued that ``\textit{people's internal relevance models are impacted by 
the relevance of the documents they initially view and that they can re-calibrate these models as they encounter
documents with more diverse relevance scores}.''

In 2018, 
as a follow up to an earlier study from 2016~\cite{damessie16},
\citet{damessie18}
compared three document ordering methods:
the aforementioned prioritisation by {\tt NTCIRPOOL},
randomised ordering,
and their own method
which presents document blocks,
where each block contains 
likely nonrelevant documents followed by 
a single pseudorelevant document.
While this third method is beyond the scope of the present study 
as it has not yet been tested in actual evaluation venues,
their results suggested that prioritisation
achieves a higher inter-assessor agreement than randomisation.
However, it should be noted that
their experiments relied on only 240 topic-document pairs:
 eight topics (4 from TREC-7, 4 from TREC-8), each with 30 pooled documents.\footnote{
 Note that ``30 pooled documents'' is not the same as depth-30.
 In general, a depth-30 pool from multiple runs would give us many more than 30 documents.
 Our depth-15 pools gave us over 200 documents per topic  (See Section~\ref{ss:constructing}).
 }
 In contrast, the present study utilises our WWW3E8 data set
with 32,375 topic-document pairs and 
 $8*32,375=259,000$ relevance labels,
 and therefore is far larger in scale.
 Note also that assessor efficiency was outside the scope of 
 the study by \citet{damessie18}.
 
 As we have mentioned in Section~\ref{s:intro},
 \citet{damessie18} pointed out that
 sorting pooled documents by document IDs
 can cluster documents by source 
 because the source IDs are often encoded as
 document ID prefixes, and that
 this can inadvertently produce small clusters of relevant documents
 for the assessors.
 For this reason, they replaced their earlier approach of 
 sorting by document IDs~\cite{damessie16}
 with randomisation.
 The present study follows suit
 and compare the RND and PRI approaches.

Also in 2018, \citet{losada18}
reported on a study on when to stop judging documents to reduce the assessment cost, under the premise that 
pooled documents are ranked by a kind of pseudorelevance.
They remark: ``{\it Although there is still room for debate,
we believe that a relevance-based ordering of assessments should not be 
an obstacle in practice.}''
Their view appears to be generally in line with the PRI approach.

Finally, the work of \citet{sakai19airs} served as a small pilot study 
for addressing our research questions \textbf{RQ1} (efficiency) and \textbf{RQ2} (inter-assessor agreement).
They utilised data from the NTCIR-14 WWW-2 task for addressing the former,
and those from the NTCIR-13 WWW-1 task for addressing the latter.
Unfortunately, however, 
their results suffered from the same bug as our original TOIS paper~\cite[corrigendum]{sakai22tois},
and therefore their results are also incorrect.

\section{Constructing the WWW3E8 Dataset}\label{s:WWW3E8}

\begin{figure}[t]

\begin{small}
\begin{verbatim}
<query>
		<qid>0001</qid>
		<content>Halloween picture</content>
		<description>Halloween is coming. You want to find some pictures about Halloween to 
		                  introduce it to your children.</description>
</query>
     :
<query>
		<qid>0101</qid>
		<content>Global military rankings</content>
		<description>You want to investigate the military powers of countries around the world.
		</description>
</query>
     :
\end{verbatim}
\caption{Topic 0001 (from the WWW-2 topic set) and 0101 (from the new WWW-3 topic set).}\label{f:topic}
\end{small}
\end{figure}

\subsection{NTCIR-15 WWW-3 English Subtask}\label{s:www3}

This section briefly describes the English subtask of the NTCIR-15 WWW-3 task~\cite{sakai20www3},
as WWW3E8 was constructed while we served as a subgroup of the organisers for this task.
WWW-3 was a traditional adhoc web search task
which offered Chinese and English subtasks.
The test topics were released in March 25, 2020;
the run submission deadline was May 31, 2020;
the task was concluded at the NTCIR-15 conference in December 2020.
The target corpus for the English subtask was clueweb12-B13.\footnote{
\url{https://lemurproject.org/clueweb12/}
}
This subtask received 37 runs from 9 teams.

The WWW-3 English subtask participants were required to process 
160 topics.
The first 80 topics (0001-0080) were constructed as the test topics at the NTCIR-14 WWW-2 task~\cite{mao19},
and the participants were given access to the WWW-2 version of the qrels file for these topics prior to run submission.
Hence it was possible for participants to tune their systems with these topics.
The other 80 topics (0101-0180) were created at WWW-3~\cite{sakai20www3}.
The common topic set size (80 for both WWW-2 and WWW-3) was determined
at the WWW-2 task~\cite{mao19}
based on \emph{topic set size design}~\cite{sakai16irj,sakai18book}:
it was estimated that 80 topics was more than sufficient
for ensuring 80\% statistical power for any $t$-test at the 5\% significance level
where the true difference between two systems is 0.10 or larger in terms 
of normalised Expected Reciprocal Rank (nERR).
nERR was
the least statistically stable measure used in the task.
We decided to conduct relevance assessments for all 160 topics from scratch,
based on the sample size considerations of \citet{sakai19airs},
although, as was mentioned earlier, their results were also affected by the same bug
that we found for our original TOIS paper.
More details on the qrels construction step will be given in Section~\ref{ss:constructing}.

The 160 topics are publicly available.\footnote{
\url{https://waseda.box.com/www2www3topics-E}
}
Thirty of the WWW-2 topics and thirty of the WWW-3 topics
originate from query logs of Sogou, a major Chinese search engine;\footnote{\url{https://www.sogou.com/}}
they were manually translated into English by the WWW organisers.
The remaining 100 topics were sampled from the AOL query log.
The WWW organisers ensured that the topic sets represent
primarily \emph{torso} queries~\cite{clarke10},
as head queries such as ``facebook.com'' are less interesting for modern web search research.
Figure~\ref{f:topic} shows Topics 0001 and 0101 as examples:
each \emph{query} (i.e., \emph{topic}) contains a \emph{content} (or \emph{title} in TREC parlance) field
and a \emph{description} field.
Both queries originate from Sogou.
The descriptions
were composed by the WWW organisers to 
back-fit the
intent and context behind the query.

%Note that the 160 topics are not part of the contribution of the present study;
%rather, our work builds on it.

\subsection{Constructing the Qrels Files}\label{ss:constructing}

%We are interested in the effect of document presentation order
%for relevance assessors
%on assessor efficiency, test collection reliablity, and reusability.
%To explore these questions,
%we decided to construct multiple versions of qrels
%for the above 160 topics, not just for the new 80 WWW-3 test topics.
%This decision was based on a sample size estimation results of Sakai and Xiao~\cite{sakai19airs}
%for comparing inter-assessor agreements across the RND and PRI conditions.
%In their experiments,
%four independent qrels files were constructed for 50 topics from the NTCIR-13 WWW-1 task~\cite{luo17}:
%two constructed based on RND pools,
%and the other two constructed based on PRI pools.
%For each of the two strategies,
%the mean per-topic inter-assessor agreement between the two qrels files 
%in terms of quadratic weighted $\kappa$~\cite{sakai19clefbook} were computed.
%According to a paired $t$-test,
%the difference between the two mean $\kappa$'s was not statistically significant
%($n=50, t=1.53, p\mbox{-}\mathrm{value}=0.134$).
%By plugging in the $t$ and $n$ values to Sakai's power analysis script 
%\texttt{future.sample.pairedt}~\cite{sakai16sigir,sakai18book}\footnote{
%Available in \url{https://waseda.box.com/SIGIR2016PACK} \ .
%},
%they found that 
%about 
%135 (150) topics are needed to achieve 70\% (75\%) statistical power
%for the comparison of the two mean $\kappa$'s.
%Since we also want the new WWW-3 test topics to satisfy
%the topic set size requirement for comparing the participating runs (See Section~\ref{s:www3}),
%we settled with a total of 160 topics, including the new 80 topics constructed at WWW-3.

Based on the recommendation from \citet{sakai19airs},
we decided to construct four versions of qrels based on RND pools,
and another four based on PRI pools,
for all of the aforementioned 160 topics. The construction procedure is described below.

Based on our budget,
we formed depth-15 pools from the 37 submitted runs.
This gave us 32,375 topic-document pairs to judge.
Hereafter, we shall refer to topic-document pairs as \emph{topicdocs} for brevity.
The average pool size is $32,375/160=202.3$.
Two versions of pool files were created for each topic:
one in which the document order is randomised (RND),
and one based on \texttt{NTCIRPOOL} (PRI).
Constructing eight independent versions of qrels (four RND-based and four PRI-based)
meant that 
a total of
$32,375*8=259,000$ labels were required.
%We constructed four RND-based pools,
%and four PRI-based tools using \texttt{NTCIRPOOL}.
%For each strategy, 
%having four independent qrels files means
%that we can obtain $4*3/2=6$ different system ranking pairs
%for a given evaluation measure;
%hence we can obtain six Kendall's $\tau$ scores.
%Similarly, 
%for the same evaluation measure,
%we can obtain $4*4=16$ Kendall's $\tau$ scores
%to compare across a RND-based ranking and a PRI-based system ranking.
%Thus we can discuss the differences in mean $\tau$'s
%within the RND condition, within the PRI condition,
%and across the two conditions,
%with sample sizes 6, 6, and 16, respectively.
%To achieve this, a total of
%$32,375*8=259,000$ labels were required.

\begin{figure}[t]
\begin{center}

\includegraphics[width=0.7\textwidth]{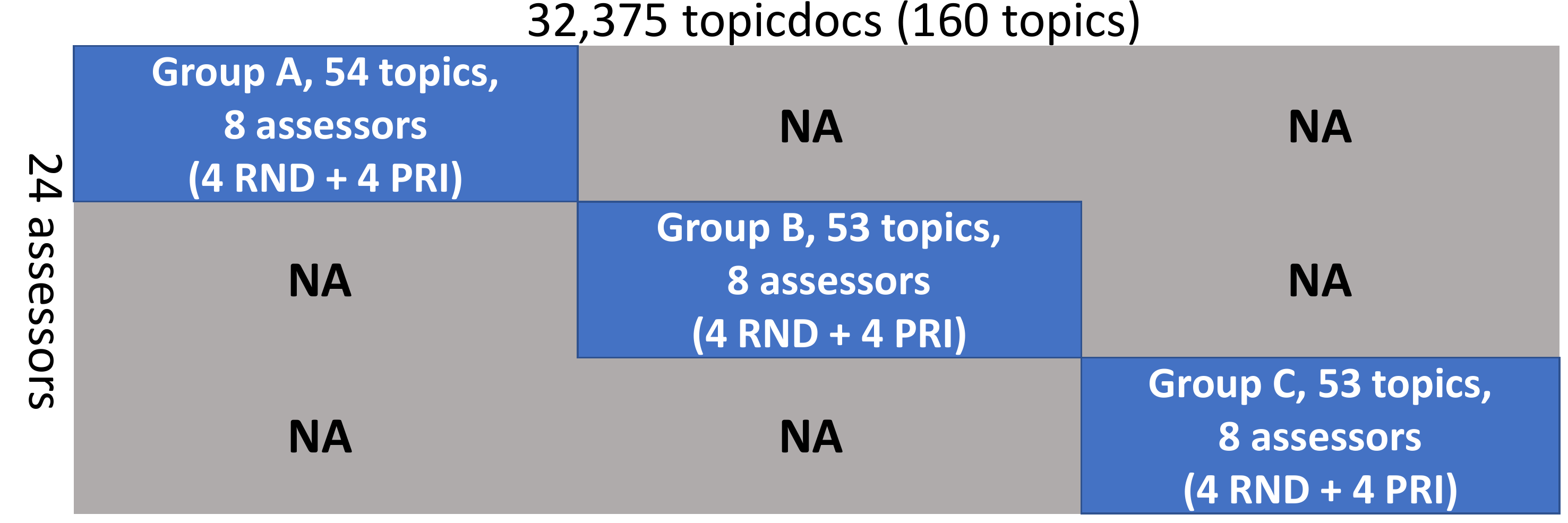}\vspace*{-2mm}
\caption{Structure of our assessor $\times$ topicdoc label matrix.}\label{f:labelmatrix}\vspace*{-5mm}

\end{center}
\end{figure}

As relevance assessors,
we hired 24 international-course (English-based programme) computer science students from our university.\footnote{
Our labelling procedure conforms to the guidelines provided by the 
Office of Research Ethics of our university.
}
Ten of them were undergraduates; the rest were master students.
We divided them at random into three groups, each consisting of eight assessors,
in order to
obtain a  24 $\times$ 32,375 relevance assessment label matrix 
that is structured as shown in Figure~\ref{f:labelmatrix}.
Group~A was given 
54 topics (11,154 topicdocs),
Group~B was given
53 topics (10,838 topicdocs),
and 
Group~C was given
the remaining 53 topics (10,383 topicdocs).
Within  each group, 
the topics and the pools (either RND or PRI) were randomly assigned,
while making sure that every assessor experienced both RND and PRI conditions in a reasonably 
balanced manner.

Based on past statistics~\cite{sakai19airs},
we told the assessors that
they are expected to spend 20 seconds on each document \emph{on average}
and that they will be paid on that basis.
This meant that the total work hours 
of each assessor in Group A-C were 62, 61, 58, respectively.
We also told them in advance that inter-assessor agreement will be checked upon job completion
for quality control.
The assessors were not aware that we were experimenting with two document ordering strategies;
they were only told that their relevant assessments will be used for web search evaluation.
The hourly pay was 1,200JPY.
Hence the total cost for constructing the label matrix
was $(62+61+58)*8*1,200=1,737,600$JPY.
The assessors were given two months (from June 15 to August 15) to complete the job;
during this period,
we sent them a reminder with each assessor's progress statistics, approximately once every two weeks.
All the assessors managed to meet the deadline. 

We instructed the assessors to go through a two-page manual first. 
The manual concisely describes the relevance assessment task,
and how to use our browser-based relevance assessment interface called PLY~\cite{luo17,sakai19clefbook}.
Each assessor was given a user account to log on to PLY,
and was allowed to work from any place where a desktop PC with internet access is available,
at any time during the above two months.
%We never physically met the assessors.

\begin{figure}[t]
\begin{center}

\includegraphics[width=\textwidth]{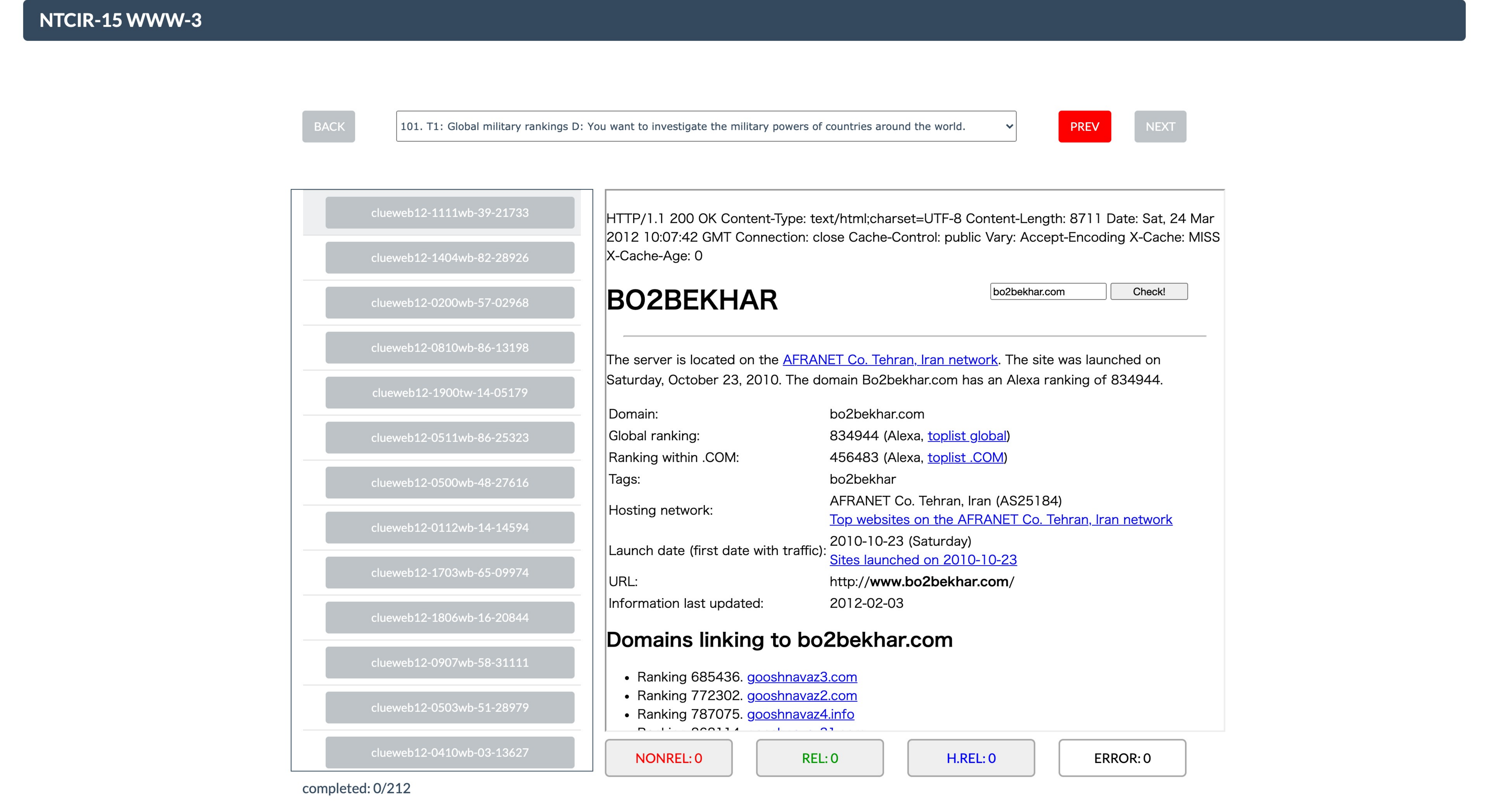}
\caption{The PLY relevance assessment interface showing a RND document list for Topic 0101.}\label{f:ply}

\end{center}
\end{figure}

Figure~\ref{f:ply} shows a screenshot of the PLY interface where the RND-based pool file for Topic 0101
has been loaded.
In principle, the assessor can judge documents in any order by clicking on a document in the left panel.
In practice, however,
they usually judge the documents from top to bottom, as judging one document automatically takes them to the next document.
As can be seen, for each document, the assessors were required to choose from four labels by clicking on a button.
These labels were defined in the aforementioned manual as follows.
\begin{description}
\item[H.REL] highly relevant - it is \textbf{likely} that the user with the information need shown will find this page  relevant.
\item[REL] relevant - it is \textbf{possible} that the user with the information need shown will find this page  relevant.
\item[NONREL] nonrelevant - it is \textbf{unlikely} that the user with the information need shown will find this page  relevant.
\item[ERROR] the right panel does not show any contents at all, even after waiting for a few seconds for the  content to load.
\end{description}
The fourth option is needed because the clueweb corpus has character encoding problems with some documents.
We thus obtained 28,144 \textbf{H.REL}, 61,512 \textbf{REL}, 
163,090 \textbf{NONREL}, and
6,254 \textbf{ERROR} labels in total.
We then treated each \textbf{H.REL} label as 2-relevant (\textbf{highly relevant}), 
each \textbf{REL} label as 1-relevant (\textbf{relevant}),
and each \textbf{NONREL} or \textbf{ERROR} label as a 0-relevant,
to form 3-point graded relevance data.

\begin{figure}[t]
\begin{center}

\includegraphics[width=\textwidth]{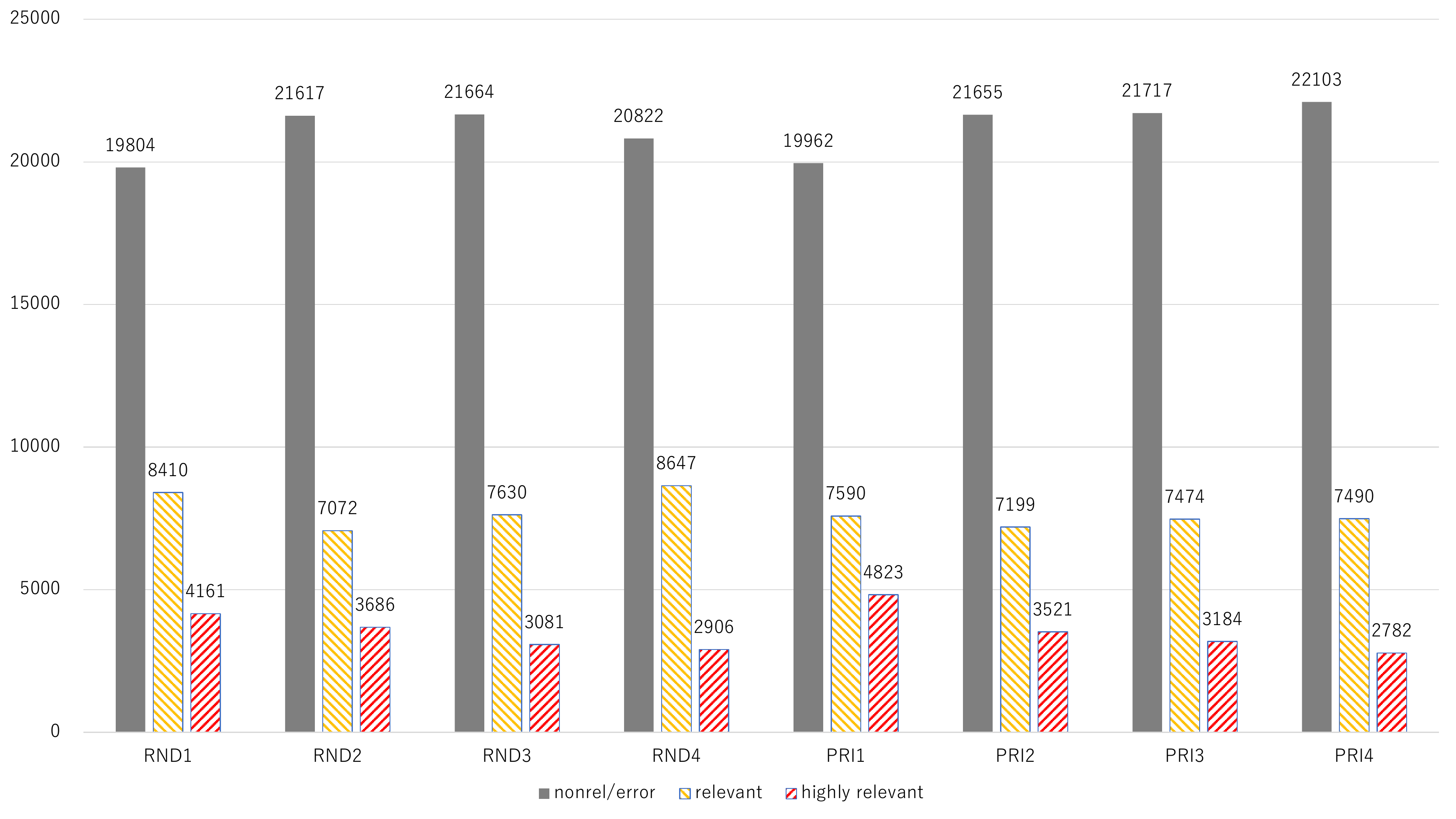}
\caption{Distribution of highly relevant (2), relevant (1), and nonrel/error (0) documents in 
each qrels file. Note that each group adds up to 32,375, i.e., the total number of topicdocs.}\label{f:rlevel-distribution}

\end{center}
\end{figure}

Figure~\ref{f:rlevel-distribution} shows the distribution of relevance labels thus obtained for each qrels file.
It can be observed that the distributions are quite similar and there are no noticeable differences
between PRI and RND.

%Figure~\ref{f:master} shows parts of the actual matrix provided in
%WWW3E8, which we call \texttt{ntcir15www2+3rawqrels.master}.
%This is the transposed matrix of what is shown in Figure~\ref{f:labelmatrix}.
%From this matrix, we created eight different qrels files
%in the simple NTCIR format\cite[p.93]{sakai19clefbook},
%which we named
%\texttt{ntcir15www2+3PRI[1-4].qrels} and\\
%\texttt{ntcir15www2+3RND[1-4].qrels}.
%Figure~\ref{f:qrels} shows parts of 
%two of these files.
%The relevance levels are prefixed with an \texttt{L}.

\subsection{Data Reliability}\label{s:reliability}

Our relevance assessors are students: they are neither topic originators
nor topic experts (i.e., \emph{bronze} assessors~\cite{bailey08}).
\citet{chouldechova13} 
report that 
query owners (i.e., topic originators) 
provide more valuable relevance assessments than non-owners for web search evaluation. 
Hence, while WWW3E8 is a large-scale relevance assessment data,
the reliability of the labels
may not be as high as those that gold assessors (i.e., topic originators)
might have provided.
To shed some light on this issue,
this section discusses data reliability  in terms of Krippendorff's $\alpha$~\cite{krippendorff18,sakai19clefbook}.

As illustrated in Figure~\ref{f:labelmatrix},
the entire label data set is a  $24 \times 32,375$ matrix
 containing 2's (\textbf{highly relevant}),
1's (\textbf{relevant}), 0's, and NA's (for Not Available).
Each topicdoc has eight labels plus 16 NA's,
since three topic subsets were handled by Groups A, B, and C, respectively.
First, from this original label matrix,
we computed Krippendorff's $\alpha$ for ordinal labels~\cite{krippendorff18,sakai19clefbook},
and obtained 
%0.288.
%This seems reasonable since, just like Cohen's $\kappa$ (which can only be used for two assessors),\footnote{
%Some researchers use Fleiss' $\kappa$ for quantifying the agreement between more than two assessors.
%However, it should be noted that Fleiss' $\kappa$ was designed for \emph{nominal} labels~\cite{fleiss71}.
%}
% Krippendorff's $\alpha$ represents agreement beyond chance.
0.425.

Next, to examine whether each assessor did a conscientious job,
we computed \emph{leave-one-out} $\alpha$ scores~\cite{sakai20tois}.
For example, we replace all labels contributed by Assessor~01  in the $24 \times 32,375$ label matrix with NA's, and then  recompute the $\alpha$.
If the $\alpha$ goes up as a result, that means that
Assessor~1 was hurting the overall reliability of the original label matrix.
%Table~\ref{t:loo} 
Table~\ref{t:loo-alpha}
shows the leave-one-out $\alpha$ scores.\footnote{
This table also correctes Table~1 of our SIGIR resource paper~\cite{sakai21www3e8}.
}
For example, the result of replacing Assessor~01's labels with NA's is shown as ``w/o A01.''
It can be observed that there are no outliers.
The results suggest that the reliability of the assessments from the 24 assessors
are comparable to one another.
%We acknowledge, however, 
%that the results do not guarantee the \emph{accuracy} of the relevance assessments.
%We only claim that WWW3E8 should serve as a useful common ground for researchers to
%compare the RND and PRI strategies within our bronze assessor setting.

\begin{table}[t]
\centering

%\caption{Krippendorff's $\alpha$ scores (for ordinal data)
%when each assessor's labels are replaced with ``NA''
%from the original 24$\times$32,375 assessor-by-topicdoc label matrix for the 160 topics.
%For example, ``w/o A01'' means ``when all labels of Assessor~01'' are replaced with an ``NA.''
%The $\alpha$ with the original matrix is 0.288.
%}\label{t:loo}
%\begin{tabular}{c|c|c|c|c|c|c|c}
%\hline
%w/o A01	&w/o A02	&w/o A03	&w/o A04	&w/o A05	&w/o A06	&w/o A07	&w/o A08\\
%\hline
%0.287		&0.287	&0.290	&0.296	&0.293	&0.290	&0.293	&0.287\\
%\hline
%\hline
%w/o A09	&w/o A10	&w/o A11	&w/o A12	&w/o A13	&w/o A14	&w/o A15	&w/o A16\\
%\hline
%0.288		&0.285	&0.286	&0.281	&0.283	&0.283	&0.281	&0.285\\
%\hline
%\hline
%w/o A17	&w/o A18	&w/o A19	&w/o A20	&w/o A21	&w/o A22	&w/o A23	&w/o A24\\
%\hline
%0.294		&0.289	&0.290	&0.294	&0.289	&0.289	&0.287	&0.289\\
%\hline
%\end{tabular}

\caption{Krippendorff's $\alpha$ scores (for ordinal data)
when each assessor's labels are replaced with ``NA''
from the original 24$\times$32,375 assessor-by-topicdoc label matrix for the 160 topics.
For example, ``w/o A01'' means ``when all labels of Assessor~01'' are replaced with an ``NA.''
The $\alpha$ with the original matrix is 
0.425.
}\label{t:loo-alpha}
\begin{tabular}{c|c|c|c|c|c|c|c}
\hline
w/o A01	&w/o A02	&w/o A03	&w/o A04	&w/o A05	&w/o A06	&w/o A07	&w/o A08\\
\hline
0.425	&0.423		&0.431		&0.432		&0.432		&0.426		&0.431		&0.422\\
\hline
\hline
w/o A09	&w/o A10	&w/o A11	&w/o A12	&w/o A13	&w/o A14	&w/o A15	&w/o A16\\
\hline
0.426	&0.424		&0.422		&0.417		&0.417		&0.420		&0.417		&0.422\\
\hline
\hline
w/o A17	&w/o A18	&w/o A19	&w/o A20	&w/o A21	&w/o A22	&w/o A23	&w/o A24\\
\hline
0.430	&0.426		&0.428		&0.432		&0.426		&0.423		&0.422		&0.427\\
\hline
\end{tabular}

\end{table}

\subsection{Runs and Score Matrices}\label{ss:runs-and}

WWW3E8 also contains the original 37 runs that contributed to the depth-15 pools 
and the topic-by-run score matrices 
for the measures used at the NTCIR-15 WWW-3 task,
created with each of the eight qrels files.
Along with the run files,
WWW8E3 also contains a file called \texttt{Eruns-1} which is a list of 36 run file names.
One run has been excluded because this run does not represent a single system:
the first 80 topics and the other 80 topics were processed by different systems~\cite{sakai20www3}.
Hence, for evaluation experiments, using this run is not recommended.

As with the official results of the NTCIR-15 WWW-3 task,
we used the \texttt{NTCIREVAL} toolkit~\cite{sakai19clefbook}\footnote{
\url{http://research.nii.ac.jp/ntcir/tools/ntcireval-en.html} (version 200626)
}
to compute all evaluation measures at the measurement depth of 10. The official measures 
are the Microsoft version of
normalised Discounted Cumulative Gain (nDCG)~\cite{sakai14promise},
the cutoff-based Q-measure~\cite{sakai14promise},
nERR (See Section~\ref{s:www3}), and intentwise Rank-Biased Utility (iRBU)~\cite{sakai20tois,sakai21ecir}.
According to the experiments by \citet{sakai19sigir,sakai20tois},
nDCG and iRBU outperformed other well-known measures in terms of agreement 
with users' SERP preferences.
%The scripts mentioned hereafter are all included in \texttt{NTCIREVAL}.

As each of the qrels files contains 3-point relevance levels (0, 1, and 2),
we let the gain values be $2^{1}-1=1$ for 1-relevant documents, and $2^{2}-1=3$ for 2-relevant documents (i.e., exponential 
gain value setting). As for the patience parameter $p$ of iRBU,
we used $p=0.99$, the default value set in \texttt{NTCIREVAL}
based on the results of \citet{sakai19sigir,sakai20tois}.

\section{RQ1: Efficiency}\label{s:efficiency}

While constructing WWW3E8, we obtained assessor activity logs from the PLY relevance assessment 
interface (See Figure~\ref{f:ply}).
Following \citet{sakai19airs},\footnote{
Both \citet{sakai19airs} and our original TOIS paper~\cite{sakai22tois}
reported ``TJ1(R)H'' (Time to \emph{judge} the first (highly) relevant document) by mistake,
although this did not substantially affect the main conclusions.
TJ1(R)H represent time to process one document,
while TF1(R)H (which were what we wanted)
generally represent time to process multiple documents. 
}
we collected the following efficiency statistics for each topic-assessor pair
(i.e., for each topic-qrels pair)
 to address \textbf{RQ1},
our assessment efficiency question.
\begin{description}
\item[TJ1D] Time to judge the first document.
\item[TF1RH] Time to find the first relevant or highly relevant document.
\item[TF1H] Time to find the first highly relevant document.
\item[ATBJ] Average time between judging two documents.
\item[NREJ] Number of times the label of a judged document is corrected to another label.
\end{description}
%For \textbf{TJ1D}, \textbf{TF1RH} and \textbf{TF1H},
%times longer than three minutes were considered outliers and were replaced with an ``NA,''
%as we cannot tell from the log whether the assessors were actually reading a document or doing something else.
%Similarly, for computing \textbf{ATBJ}, times longer than three minutes were excluded when computing the average.\footnote{
%We decided to adopt this 3-minute threshold from Sakai and Xiao~\cite{sakai19airs},
%after observing that 
%the \textbf{ATBJ} statistics in our experiments are similar to theirs (about 14 seconds as shown in Table~\ref{t:efficiency}).
%}
Note that \textbf{ATBJ} is the most direct measure of assessor efficiency;
when computing each \textbf{ATBJ}, times longer than three minutes were considered outliers and were excluded when averaging.
Similarly, for \textbf{TJ1D}, times longer than three minutes were replaced with an ``NA.''
As for \textbf{TF1RH} and \textbf{TF1H}, times longer than \emph{30 minutes} were replaced with an ``NA,''
as these statistics generally represent times to process \emph{multiple} documents.

Table~\ref{t:efficiency} shows, for each qrels file (PRI1 through RND4),
our five efficiency criteria averaged across the topics.
Note that the sample sizes are much smaller than 160 for \textbf{TJ1D}, \textbf{TF1RH}, and \textbf{TF1H}
because we removed every topic that resulted in an ``NA'' for 
at least one version of the qrels.
For each efficiency criterion,
as we have eight mean scores to compare,
we conducted a paired Tukey HSD test~\cite{sakai18book} at the 5\% significance level.

\begin{table}[t]
\centering

\begin{small}

\caption{Efficiency comparison for the eight qrels files (PRI1 through RND4) for the 160 topics.
\textbf{TJ1D}, \textbf{TF1RH}, \textbf{TF1H}, and \textbf{TF1H} are in seconds;
\textbf{NREJ} represents the number of label corrections.
For each efficiency criterion, a paired Tukey HSD test at the 5\% significance level was conducted to compare every pair of means.
All statistically significant differences are indicated in the table with the $p$-value ($p$) and the effect size $\textit{ES}$,
and a unique symbol for each pair of qrels file.
$V_{\mathrm{E2}}$ denotes the two-way ANOVA residual variance for computing the effect sizes~\cite{sakai18book}.
In each row,
the largest and smallest among the eight values are shown in \textbf{bold} and \textit{italics}, respectively.
}\label{t:efficiency}
\begin{tabular}{c|r||c|c|c|c||c|c|c|c||c}
\hline
Criterion 		&$n$	&RND1					&RND2			&RND3	&RND4				&PRI1	&PRI2	&PRI3				&PRI4			&$V_{\mathrm{E2}}$\\
\hline
\textbf{TJ1D} 	&44		&34.3					&\textbf{41.2}			&39.7	&38.8				&34.3	&35.7	&37.1				&\textit{28.1}			&1200.5\\
\textbf{TF1RH}	&82		&\textit{85.3} 		&147.6 						&163.0 	&\textbf{206.5} 									&108.0 						&112.2 	&89.5 						&175.2	&55844\\
		&		&$\ast (p=$		&							&		&$\ast\clubsuit$									&							&		&$\clubsuit (p=$				&		&\\
		&		&0.024,			&							&		&												&							&		&0.034,						&		&\\
		&		&$\textit{ES}=	$					&		&												&							&		&							&$\textit{ES}=$	&&\\
		&		&0.513)			&							&		&												&							&		&0.495)						&		&\\
\textbf{TF1H}	&32		&261.0 			&218.1 						&263.9 	&\textbf{488.7} 									&179.2 						&190.2 	&157.3 						&\textit{129.9}	&89460\\
		&		&				&$\triangle (p=$				&		&$\triangle\heartsuit\diamondsuit\clubsuit\spadesuit$	&$\heartsuit (p=$				&$\diamondsuit (p=$			&$\clubsuit (p=$	&$\spadesuit (p=$	&\\
		&		&				&0.0087,						&		&												&0.0013,						&0.0023,						&0.00039,		&0.000081,	&\\
		&		&				&$\textit{ES}=$	&		&												&$\textit{ES}=$	&$\textit{ES}=$	&$\textit{ES}=$&$\textit{ES}=$	&\\
		&		&				&0.905)						&		&												&1.03)						&0.998)						&1.11)	&1.20)	&\\
\textbf{ATBJ} 	&160	&\textit{13.1}$\heartsuit$			&\textbf{15.8}$\heartsuit$	&14.5	&14.8				&14.5	&14.3	&14.7				&14.7			&56.53\\
				&		&($p=$					&				&		&					&		&		&					&				&\\
				&		&0.028)					&				&		&					&		&		&					&				&\\
				&		&($\textit{ES}=$				&				&		&					&		&		&					&				&\\
				&		&0.361)					&				&		&					&		&		&					&				&\\
\textbf{NREJ}		&160	&\textbf{8.41}$\sharp\circ$	&6.51			&6.11	&3.99$\circ$		&6.79	&7.31	&5.99				&\textit{3.82}$\sharp$	&107.4\\
				&		&						&				&		&($p=$				&		&		&					&($p=$			&\\
				&		&						&				&		&0.0036)				&		&		&					&0.0020)		&\\
				&		&						&				&		&($\textit{ES}=$			&		&		&					&($\textit{ES}=$		&\\
				&		&						&				&		&0.427)				&		&		&					&0.443)			&\\
\hline
\end{tabular}

%\caption{Achieved statistical powers if paired $t$-tests are used for the largest differences shown in Table~\ref{t:efficiency},
%with sample sizes required for achieving 70\% power.
%Sakai's R script \texttt{future.sample.pairedt}~\cite{sakai16sigir,sakai18book} was used for the analysis.
%}\label{t:efficiency-achieved}
%\begin{small}
%\begin{tabular}{c|r|c|c|c|r}
%\hline
%Criterion			&actual $n$	&largest difference		&$t$ statistic		&achieved power	&$n$ for 70\% power\\
%\hline
%\textbf{TJ1D}		&44			&RND2 $>$ PRI4		&2.42			&65.6\%			&49\\
%\textbf{TF1RH}	&72			&PRI2 $>$ RND1		&1.33			&26.0\%			&253\\
%\textbf{TF1H}		&59			&PRI4 $>$ RND1		&3.01			&84.1\%			&43\\
%\textbf{ATBJ}		&160		&RND2 $>$ RND1		&4.00			&97.8\%			&64\\
%\textbf{NREJ}		&160		&RND1 $>$ PRI4		&3.85			&96.9\%			&69\\
%\hline
%\end{tabular}

\caption{Achieved statistical powers if paired $t$-tests are used for the largest differences shown in Table~\ref{t:efficiency},
with sample sizes required for achieving 70\% power.
Sakai's R script \texttt{future.sample.pairedt}~\cite{sakai16sigir,sakai18book} was used for the analysis.
}\label{t:efficiency-achieved}
\begin{tabular}{c|r|c|c|c|r}
\hline
Criterion			&actual $n$	&largest difference		&$t$ statistic		&achieved power	&$n$ for 70\% power\\
\hline
\textbf{TJ1D}		&44			&RND2 $>$ PRI4		&2.42			&65.6\%			&49\\
\textbf{TF1RH}	&82			&RND4 $>$ RND1		&3.54			&93.8\%			&43\\
\textbf{TF1H}		&32			&RND4 $>$ PRI4		&4.21			&98.3\%			&14\\
\textbf{ATBJ}		&160		&RND2 $>$ RND1		&4.00			&97.8\%			&64\\
\textbf{NREJ}		&160		&RND1 $>$ PRI4		&3.85			&96.9\%			&69\\
\hline
\end{tabular}

\end{small}
\end{table}

Our observation for each efficiency criterion is as follows.
\begin{description}
\item[TJ1D] While RND2 is the least efficient (41.2 seconds) and PRI4 is the most efficient (28.1 seconds) on average,
no statistically significant differences are observed.
From the table,
we can work out
the effect size (i.e., standardised mean difference)~\cite{sakai18book} for this largest difference as $\textit{ES}=(41.2-28.1)/\sqrt{1200.5}=0.378$, which is very small and probably not practically significant either.
\item[TF1RH] As indicated by the $\clubsuit$'s,
PRI3 is statistically significantly more efficient than RND4 ($p=0.034, \textit{ES}=0.495$).
However, we cannot conclude that PRI is generally more efficient than RND,
as RND1 is actually the most efficient on average;
RND1 also 
 statistically significantly outperforms RND4 with a similar effect size ($p=0.024, \textit{ES}=0.513$).
\item[TF1H] The means for PRI are all smaller than those for RND.
In particular,
RND4 is statistically significantly less efficient than any of the four PRI qrels files,
with an effect size of around one.
(RND2 is also statistically significantly more efficient than RND4).
From these results, it is possible that
PRI tends to enable the assessor to find the first highly relevant document in the pool
slightly more quickly compared to RND.
\item[ATBJ] The only statistically significant difference observed is between RND1 and RND2,
and the effect size is small (0.361) even for this pair.
Hence, we conclude that
there is not substantial difference between RND and PRI in terms of time spent for judging each document.
\item[NREJ] While RND1 makes corrections statistically significantly more often 
than RND4 ($p=0.0036, \textit{ES}=0.417$) and PRI4 ($p=0.0020, \textit{ES}=0.443$),
there is no clear trend regarding the RND vs. PRI comparisons.
\end{description}
In summary,
our assessor efficiency results show that
\emph{there is no substantial difference between RND and PRI
in terms of time spent for judging each document,
although PRI may enable
faster identification of the first highly relevant document in the pool.}

%As was discussed in Section~\ref{ss:airs2019},
%we aimed for 70\% statistical power (when paired $t$-tests are used)
%based on the sample size considerations of Sakai and Xiao~\cite{sakai19airs}.
When deciding on the size of the WWW3E8 data set, 
we aimed for 70\% statistical power with paired $t$-tests~\cite{sakai19airs}.
Table~\ref{t:efficiency-achieved} examines the actual achieved statistical powers
for the largest differences observed in Table~\ref{t:efficiency},
using the R script \texttt{future.sample.pairedt} from \citet{sakai16sigir,sakai18book}.\footnote{
Available from \url{https://waseda.box.com/SIGIR2016PACK} .
}
%For example,
%if we conduct 
%a paired $t$-test with RND2 and PRI4 for the \textbf{TJ1D} row of Table~\ref{t:efficiency},\footnote{
%Note that we do not actually discuss this $t$-test: instead, Table~\ref{t:efficiency} uses the Tukey HSD test (i.e., a multiple comparison procedure) that utilises all eight versions of qrels.
%}
%the achieved power is about 65.6\%, which is close to what we originally aimed for.
%It can be observed that, with the exception of \textbf{TF1RH},
%we have managed to achieve high statistical powers.
%Finally, 
%while our experiment is underpowered for \textbf{TF1RH},
%whether this deserves further investigation with a larger sample is questionable,
%given the above discussions of small effect sizes and assessor variance.
%Recall that the largest difference in terms of \textbf{TF1RH} in Table~\ref{t:efficiency}
%had an effect size of only 0.211; 
%in Sakai and Xiao~\cite{sakai19airs},
%the corresponding Glass's $\Delta$~\cite{sakai18book} was 0.305.
It can be observed that, except for \textbf{TJ1D},
each efficiency criterion in our experiments achieved a statistical power of over 90\%.
The analysis suggests that our sample sizes are sufficiently large for discussing these efficiency statistics.

\section{RQ2: Inter-Assessor Agreement}\label{s:agreement}

\begin{table}[t]
%\centering

%\vspace*{5mm}\caption{Overall inter-assessor agreement in terms of 
%Krippendorff's $\alpha$  (for ordinal data) based on the 24 $\times$ 32,375 label matrices for the 160 topics.
%The original matrix contains 8 labels per topicdoc (4 based on RND, 4 based on PRI);
%the RND and PRI matrices each contain 4 labels per topicdoc. All other cells are stuffed with ``NA.''
%}\label{t:overall}
%\begin{scriptsize}
%\begin{tabular}{c|c|c}
%\hline
%All		&RND	&PRI\\
%\hline
%0.288		&0.433	&0.423\\
%\hline
%\end{tabular}

\caption{Overall inter-assessor agreement in terms of 
Krippendorff's $\alpha$  (for ordinal data) based on the 24 $\times$ 32,375 label matrices for the 160 topics.
The original matrix contains 8 labels per topicdoc (4 based on RND, 4 based on PRI);
the RND and PRI matrices each contain 4 labels per topicdoc. All other cells are stuffed with ``NA.''
}\label{t:overall-alpha}
\begin{tabular}{c|c|c}
\hline
All		&RND	&PRI\\
\hline
0.425	&0.433	&0.423\\
\hline
\end{tabular}

%\end{scriptsize}
%\end{table}

%\begin{table}[t]
%\centering

%\caption{Mean per-topic
%Krippendorff's $\alpha$  (for ordinal data) averaged over the 160 topics.
%Each per-topic matrix contains 4 labels (either based on RND or PRI) per document.
%All other cells are stuffed with ``NA.''
%The mean $\alpha$ when all 8 labels are included in the matrix is 0.125.
%A paired $t$-test at the 5\% significance level was conducted.
%Glass's $\Delta$~\cite{sakai18book} is based on the standard deviation from the RND data.
%}\label{t:mean-alpha}
%\begin{small}
%\begin{tabular}{c|c|c|c|c|l|c|c}
%\hline
%$n$	&RND		&PRI		&$t$ statistic	&$p$-value	&Glass's $\Delta$	&Achieved power	&$n$ for 70\% power\\
%\hline
%160	&0.293		&0.279	&0.949		&0.344		&0.0859			&15.7\% &$n=1,098$\\
%\hline
%\end{tabular}

\caption{Mean per-topic
Krippendorff's $\alpha$  (for ordinal data) averaged over the 160 topics.
Each per-topic matrix contains 4 labels (either based on RND or PRI) per document.
All other cells are stuffed with ``NA.''
The mean $\alpha$ when all 8 labels are included in the matrix is 
0.288.
A paired $t$-test at the 5\% significance level was conducted.
Glass's $\Delta$~\cite{sakai18book} is based on the standard deviation from the RND data.
Sakai's R script \texttt{future.sample.pairedt}~\cite{sakai16sigir,sakai18book} was used for the analysis.
}\label{t:mean-alpha}
\begin{tabular}{c|c|c|c|c|l|c|c}
\hline
$n$	&RND		&PRI		&$t$ statistic	&$p$-value	&Glass's $\Delta$	&Achieved power	&$n$ for 70\% power\\
\hline
160	&0.293		&0.279	&0.949		&0.344		&0.0859			&15.7\% &$n=1,098$\\
\hline
\end{tabular}

%\vspace*{5mm}\caption{Verifying the achieved statistical power in terms of mean kappa.
%}\label{t:kappa-achieved}
%\begin{small}
%\begin{tabular}{c|c|c|c|c|c|c}
%\hline
%$n$	&RND1 vs.		&RND2 vs. 	&$t$ statistic	&$p$-value	&Achieved power	&$n$ for 70\% power\\
%	&RND4		&RND3	&			&			&				&\\
%\hline
%160	&0.175				&0.141			&3.33		&0.00108		&91.1\%			&91\\
%\hline
%\end{tabular}

\caption{Verifying the achieved statistical power in terms of mean quadratic weighted $\kappa$. The largest mean $\kappa$ (for RND1 vs RND4)
and the smallest (for RND1 and PRI1) are used for this analysis.}\label{t:kappa-achieved}
\begin{tabular}{c|c|c|c|c|c|c}
\hline
$n$	&RND1 vs.	&RND1 vs. 	&$t$ statistic	&$p$-value	&Achieved power	&$n$ for 70\% power\\
	&RND4		&PRI1		&			&			&				&\\	
\hline
160		&0.175	&0.130		&4.57		&0.0000097	&99.5\%			&50\\
\hline
\end{tabular}

%\end{small}
\end{table}

We now utilise the WWW3E8 data set to address
\textbf{RQ2} (Which document ordering strategy
enables higher inter-assessor agreements?).
We quantify the inter-assessor agreements under
RND and PRI conditions
using Krippendorff's $\alpha$
for ordinal classes~\cite{krippendorff18,sakai19clefbook}.
For example, to quantify the inter-assessor agreement
under the RND condition,
all labels in the original WWW3E8 matrix
 that were obtained under the PRI condition can be replaced with NA's
 and then the $\alpha$ can be recomputed,
so that each topicdoc has only four labels instead of eight.
Table~\ref{t:overall-alpha} shows the results.
It can be observed that the 
three $\alpha$ scores (using all eight labels
vs. using only the four RND labels vs. using only the four PRI labels)
are very similar.
%It can be observed that while the $\alpha$ scores for RND and PRI
%are very similar, they are much higher than 
%the $\alpha$ score for the original matrix (0.288).
%That is, while the labels \emph{within} each document ordering strategy are similar to each other,
%the labels \emph{across} the two strategies differ substantially.
%It is clear that \emph{the document ordering strategy substantially affects 
%which documents are judged (highly) relevant.}
%We shall provide an explanation for this in
%Section~\ref{ss:order}.

The above analysis computed a single $\alpha$ score for the entire matrix.
In contrast,
Table~\ref{t:mean-alpha} compares the inter-assessor agreement under the RND and PRI conditions
based on mean \emph{per-topic} $\alpha$ scores, averaged over the 160 topics.
According to a paired $t$-test, the difference 
between the RND and PRI conditions is not statistically significant.
More importantly,
the effect size (Glass's $\Delta$, a form of standardised mean difference~\cite{sakai18book})
in terms of  $\alpha$ is very small ($\Delta=0.0859$),
and power analysis asks for over 1,000 topics to achieve 70\% statistical power for such a small effect size.\footnote{
Sakai's tool \texttt{future.sample.pairedt}~\cite{sakai16sigir}
was used for the power analysis.
} 
From these results, we conclude that
\emph{The effect of document ordering strategy on inter-assessor agreement is negligible.}
 
% In Section~\ref{ss:airs2019}, we discussed the sample size considerations of Sakai and Xiao~\cite{sakai19airs},
% who recommended having 160 topics for achieving over 70\% statistical power for mean kappas.
% Table~\ref{t:kappa-achieved} shows the achieved statistical power 
% for the largest difference among mean weighted kappas of all pairs of qrels,
% namely, the difference between the kappa for RND1 and RND4, 
% and that for RND2 and RND3.
% It can be observed that the achieved power is over 91\%.
Our decision to use as many as 160 topics was based 
on sample size considerations from \citet{sakai19airs}
where the aim was to achieve over 70\% statistical power for mean $\kappa$'s (rather than Krippendorff's $\alpha$).
Hence, for completeness, 
Table~\ref{t:kappa-achieved} shows the achieved statistical power 
for the largest difference among mean weighted $\kappa$'s of all pairs of qrels.
The result shows that our experiment has a very high statistical power.
 
% We also observe from Table~\ref{t:kappa-achieved} that 
% the pairwise agreements in terms of kappa are lower compared to Sakai and Xiao~\cite[Table~3]{sakai19airs}:
% their kappa values lay between 0.247 (PRI1 vs. PRI2) and 0.341 (RND2 vs. PRI2).
% This suggests that there is considerable noise in the relevance assessments of the WWW3E8 data set.
% Muraoka, Zeng, and Sakai~\cite{muraoka20} manually examined a small number of documents
% as a participating team of the NTCIR-15 WWW-3 task and also 
% reported on some possible noise in the relevance assessments for the WWW-3 topics.
% Evaluating the \emph{accuracy} of relevance assessments require \emph{gold} relevance assessments~\cite{bailey08},
% and is therefore left for future work (See Section~\ref{s:conclusions}).

\section{RQ3: System Ranking Agreement}\label{s:system-ranking}

% 8 topics lost by RND-PRI, another 5 lost by LOTO. Use 160-13=147 topics in RQ3 and RQ4.
% todo.

\subsection{System Ranking $\tau$ Results}\label{ss:system-tau}

Using the eight qrels files (PRI1 through RND4) available in WWW3E8,
we now address \textbf{RQ3} (Which strategy enables more stable system rankings across different versions of qrels files?).
More specifically,
using each qrels file,
we rank the 36 runs submitted to the NTCIR-15 WWW-3 task~\cite{sakai20www3}
with the official measures used in the task;
recall that one run was excluded as described in Section~\ref{ss:runs-and}.
We then 
quantify the system ranking similarity with Kendall's $\tau$~\cite{sakai14promise}.
As we have mentioned in Section~\ref{ss:runs-and},
the official measures used in the WWW-3 task are
nDCG,
Q-measure, 
nERR,
and 
iRBU.
These were computed using 
\texttt{NTCIREVAL}
with an exponential gain value setting:
3 for \textbf{highly relevant} and 1 for \textbf{relevant}.

In this section as well as Section~\ref{s:newsystems}, 
evaluation measure scores are averaged over 147 topics rather than 160 topics for the following reasons.
(1)~There are a small number of cases where the qrels files (PRI1 through RND4)
do not have any relevant documents. 
The union of such topics across all eight qrels files amounts to
8 topics (0012, 0024, 0026, 0044, 0063, 0147, 0169, 0174).
(2)~In Section~\ref{s:newsystems} where we discuss the subsets of the above qrels files 
to address \textbf{RQ4} (robustness to new systems),
we lose 5 topics in addition for the same reason (0027, 0060, 0132, 0153, 0179).
Hence, in order to evaluate the runs using a common topic set for every experimental condition,
we average the evaluation measures over $160-8-5=147$ topics.
We have a total of 29,522 topicdocs from the depth-15 pools for these topics.

Table~\ref{t:inter-qrels-tau} shows the results of comparing all pairs of qrels versions in terms of $\tau$.
%Correlation strengths are visualised in color ($\tau > 0.8, \textcolor{red}{0.5 < \tau \leq 0.8}, \textit{\tau \leq 0.5}$).
%The trends are similar across all four evaluation measures, and are very clear. More specifically:
%\begin{itemize}
%\item The four PRI-based qrels files produce very similar system rankings ($\tau \geq 0.841$);
%\item The four RND-based qrels files produce moderately similar system rankings ($0.495 \leq \tau \leq 0.781$);
%\item The RND-based rankings and the PRI-based ones are substantially different ($\tau \leq 0.508$).
%\end{itemize}
It can be observed that, for each evaluation measure, the system rankings
according to the eight qrels files are generally similar.
The minimum $\tau$ observed in the table is  0.733 for RND2 vs. PRI4 with nERR (95\%CI[0.608, 0.822], $n=36$),
while the maximum is 0.944 for RND3 vs. RND4 with nDCG (95\%CI[0.913, 0.964], $n=36$).

The above three levels of system ranking agreement can be examined  more closely as follows.
From Table~\ref{t:inter-qrels-tau},
we can compute, for each evaluation measure,
 a mean $\tau$ that represents 
the agreement within the RND condition by averaging the six values that compare two RND-based rankings.
Similarly, we can obtain a mean $\tau$ within the PRI condition by averaging the six values that compare two PRI-based rankings.
Finally, we can obtain a mean $\tau$ across the two conditions by averaging the $4*4=16$ values that
compare a RND-based ranking and a PRI-based ranking.
To discuss the differences in means for these three cases,
we can apply a Tukey HSD test for unpaired data at the 5\% significance level~\cite{sakai18book}.

Table~\ref{t:tau-significance} shows the results of the unpaired Tukey HSD test for each evaluation measure.
%Again, the results are similar for all four measures:
%the ``Mean $\tau$'' columns show that
%system rankings within the PRI condition are very similar,
%those within the RND condition are less so,
%and that those across the two conditions are 
%substantially different.
%As the ``$p$-value'' columns show,
%all of these differences in means are statistically highly significant.
%The ``$V_{E1}$'' column shows the residual variance from one-way ANOVA for computing the effect sizes,
%since we are dealing with unpaired data here~\cite{sakai18book}.
%For example, 
%the difference in mean nDCG between the within-RND condition (RND-RND) 
%and the within-PRI condition (PRI-PRI) is 
%$0.917-0.741=0.176$;
%therefore the effect size can be computed as 
%$0.176/\sqrt{0.000831}=6.12$.
%That is, the two means are about six standard deviations apart.
%We conclude that \emph{
%different system rankings under the PRI condition are 
%substantially more similar to one another
%than those under the RND condition.}
%Moreover, as we have observed in Table~\ref{t:inter-qrels-tau},
%\emph{PRI-based rankings and RND-based rankings are substantially different from each other.}
The ``Mean $\tau$'' columns show that
on average, 
the PRI-PRI system ranking agreements are the highest 
while the RND-PRI agreements are the lowest.
On the other hand,
the ``$p$-value'' columns show that,
while the RND-PRI agreements are statistically significantly and substantially
lower than the RND-RND and PRI-PRI agreements,
the differences between the RND-RND and PRI-PRI agreements are not
statistically significant.
The largest effect size (standardimised mean difference) observed for the 
latter comparison is $\textit{ES}_{\mathrm{E1}}=(0.8975-0.8518)/\sqrt{0.00115}=1.35$ for nERR.
In short, the PRI-PRI agreements tend to be higher than
the RND-RND ones, but only slightly.

\begin{table}[t]
\centering

\begin{scriptsize}

\caption{System ranking agreement as measured by Kendall's $\tau$ between system ranking pairs ($n=36$ runs)
according to the four official measures of the NTCIR-15 WWW-3 task, averaged over 147 topics.
To visualise the general trend, 
$\tau$'s larger than 0.8 are shown in \textbf{bold}, and those smaller than 0.5 are shown in \textit{italics}.
}\label{t:inter-qrels-tau}

\begin{tabular}{c|c|c|c|c|c|c|c}
\hline
(a) nDCG &RND2		&RND3		&RND4		&PRI1		&PRI2		&PRI3		&PRI4\\
\hline
RND1	&0.860		&0.876		&0.900		&0.889		&0.908		&0.857		&0.879\\
RND2	& -			&0.895		&0.878		&0.800		&0.832		&0.800		&0.784\\
RND3	& -			& -			&0.944		&0.835		&0.848		&0.829		&0.832\\
RND4	& -			& -			& -			&0.840		&0.859		&0.833		&0.843\\
PRI1	& -			& -			& -			& -			&0.905		&0.924		&0.927\\
PRI2	& -			& -			& -			& -			& -			&0.911		&0.908\\
PRI3	& -			& -			& -			& -			& -			& -			&0.927\\
\hline
(b) Q 	&RND2		&RND3		&RND4		&PRI1		&PRI2		&PRI3		&PRI4\\
\hline
RND1	&0.857		&0.875		&0.894		&0.892		&0.922		&0.886		&0.851\\
RND2	& -			&0.862		&0.811		&0.775		&0.827		&0.819		&0.746\\
RND3	& -			& -			&0.886		&0.849		&0.848		&0.830		&0.814\\
RND4	& -			& -			& -			&0.862		&0.860		&0.840		&0.830\\
PRI1	& -			& -			& -			& -			&0.884		&0.879		&0.902\\
PRI2	& -			& -			& -			& -			& -			&0.890		&0.843\\
PRI3	& -			& -			& -			& -			& -			& -			&0.883\\
\hline
(c) nERR &RND2		&RND3		&RND4		&PRI1		&PRI2		&PRI3		&PRI4\\
\hline
RND1	&0.838		&0.841		&0.857		&0.854		&0.843		&0.816		&0.819\\
RND2	& -			&0.844		&0.829		&0.756		&0.773		&0.743		&0.733\\
RND3	& -			& -			&0.902		&0.797		&0.802		&0.759		&0.781\\
RND4	& -			& -			& -			&0.857		&0.843		&0.813		&0.841\\
PRI1	& -			& -			& -			& -			&0.884		&0.886		&0.895\\
PRI2	& -			& -			& -			& -			& -			&0.906		&0.906\\
PRI3	& -			& -			& -			& -			& -			& -			&0.908\\
\hline
(d) iRBU &RND2		&RND3		&RND4		&PRI1		&PRI2		&PRI3		&PRI4\\
\hline
RND1	&0.830		&0.808		&0.837		&0.852		&0.792		&0.849		&0.811\\
RND2	& -			&0.895		&0.883		&0.790		&0.810		&0.829		&0.806\\
RND3	& -			& -			&0.876		&0.762		&0.762		&0.794		&0.797\\
RND4	& -			& -			& -			&0.835		&0.825		&0.860		&0.838\\
PRI1	& -			& -			& -			& -			&0.848		&0.886		&0.902\\
PRI2	& -			& -			& -			& -			& -			&0.873		&0.889\\
PRI3	& -			& -			& -			& -			& -			& -			&0.883\\
\hline
\end{tabular}

%\end{small}
%\end{table}

%\begin{table}[t]
%\centering

%\caption{Comparison of mean system ranking $\tau$'s based on the $\tau$'s shown in Table~\ref{t:inter-qrels-tau}.
%Results of
%Tukey HSD tests for unpaired data with sample sizes 6, 6, 16 are shown.
%The effect sizes are standardised mean differences based on the 
%one-way ANOVA
%residual variance $V_{E1}$~\cite{sakai18book}.
%}\label{t:tau-significance}

%\begin{small}
%\begin{tabular}{c|c|c|c|c|c|c|l}
%\hline
%Measure	&\multicolumn{3}{|c|}{Mean $\tau$ (sample size)}	&Residual	&\multicolumn{3}{|c}{$p$-value (effect size)}\\
%		&RND-RND		&PRI-PRI		&RND-PRI		&variance	&RND-RND vs	&PRI-PRI vs	&PRI-PRI vs\\
%		&($n_{1}=6$)	&($n_{1}=6$)	&($n_{1}=16$)	&$V_{E1}$	&RND-PRI		&RND-PRI		&RND-RND\\
%\hline
%nDCG	&0.741		&0.917		&0.348		&0.000831	&$\approx 0$ (13.6)	&$\approx 0$ (19.8)	&$\approx 0$ (6.12)\\
%Q		&0.716		&0.880		&0.369		&0.00153		&$\approx 0$ (8.88)	&$\approx 0$ (13.0)	&$\approx 0$ (4.17)\\
%nERR	&0.635		&0.898		&0.280		&0.00172		&$\approx 0$ (8.57)	&$\approx 0$ (14.9)	&$\approx 0$ (6.34)\\
%iRBU	&0.592		&0.880		&0.419		&0.00308		&$\approx 0$ (3.12)	&$\approx 0$ (8.32)	&$\approx 0$ (5.20)\\
%\hline
%\end{tabular}

\caption{Comparison of mean system ranking $\tau$'s based on the $\tau$'s shown in Table~\ref{t:inter-qrels-tau}.
with
Tukey HSD tests for unpaired data.
Statistical significance at $\alpha=0.05$ is indicated in \textbf{bold}.
The effect sizes are standardised mean differences based on the 
one-way ANOVA
residual variance $V_{\mathrm{E1}}$~\cite{sakai18book}.
}\label{t:tau-significance}
\begin{tabular}{c|c|c|c|c|c|c|l}
\hline
Measure	&\multicolumn{3}{|c|}{Mean $\tau$ (sample size)}	&Residual	&\multicolumn{3}{|c}{$p$-value and effect size}\\
		&RND-RND		&PRI-PRI		&RND-PRI		&variance	&RND-RND vs	&PRI-PRI vs	&PRI-PRI vs\\
		&($n=6$)	&($n=6$)	&($n=16$)	&$V_{\mathrm{E1}}$	&RND-PRI		&RND-PRI		&RND-RND\\
\hline
nDCG	&0.8922		&0.9170		&0.8418		&0.000826	&$\mathbf{p < 0.004}$ 	&$\mathbf{p < 0.00003}$ 		&$p=0.31$ \\
		&			&			&			&			&\textit{ES}=1.75		&\textit{ES}=2.62		&\textit{ES}=0.86\\
Q		&0.8642		&0.8802		&0.8407		&0.00134		&$p = 0.39$ 			&$p=0.082$ 			&$p=0.73$ \\
		&			&			&			&			&\textit{ES}=0.64		&\textit{ES}=1.08		&\textit{ES}=0.44\\
nERR	&0.8518		&0.8975		&0.8019		&0.00115		&$\mathbf{p=0.013}$ 	&$\mathbf{p < 0.00002}$ 		&$p=0.070$ \\
		&			&			&			&			&\textit{ES}=1.47		&\textit{ES}=2.82		&\textit{ES}=1.35\\
iRBU	&0.8548		&0.8802		&0.8132		&0.000842	&$\mathbf{p=0.016}$ 	&$\mathbf{p < 0.0002}$ 		&$p=0.30$\\
		&			&			&			&			&\textit{ES}=1.43		&\textit{ES}=2.31		&\textit{ES}=0.87\\
\hline
\end{tabular}

\end{scriptsize}

\end{table}

%\subsection{Why Does PRI Produce Similar Rankings?}\label{ss:order}

\subsection{Why Do PRI-PRI System Ranking Agreements Tend To Be Relatively High? }\label{ss:order}

\begin{figure}[t]
\begin{center}

%\includegraphics[width=0.9\textwidth]{RND.rlevel1+2.pdf}
%\includegraphics[width=0.9\textwidth]{PRI.rlevel1+2.pdf}
%\caption{Document presentation order vs. counts of relevance labels (relevant, highly relevant) 
%based on the main experiment with the 160 topics.}\label{f:order-sum}

\includegraphics[width=0.9\textwidth]{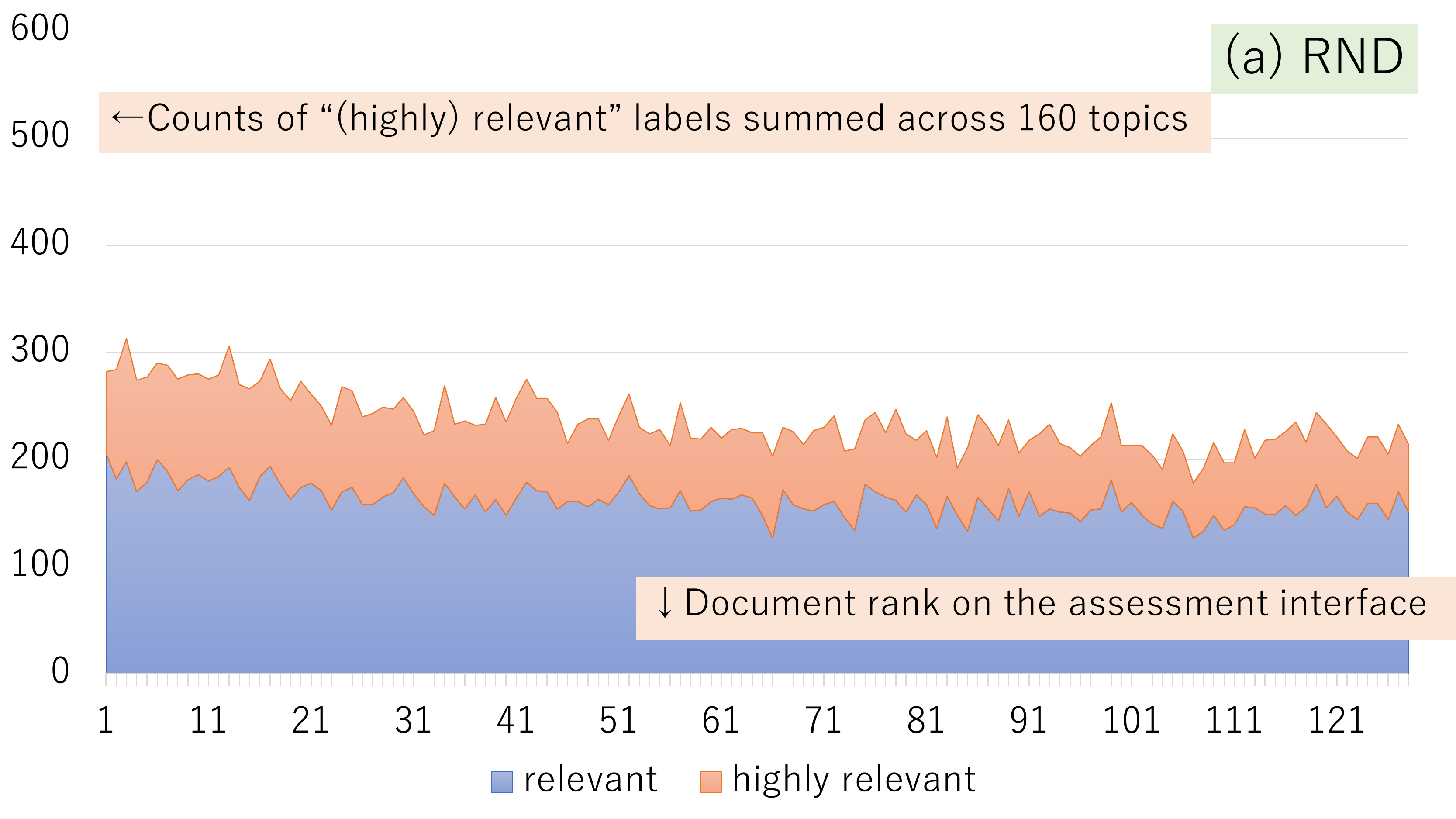}
\includegraphics[width=0.9\textwidth]{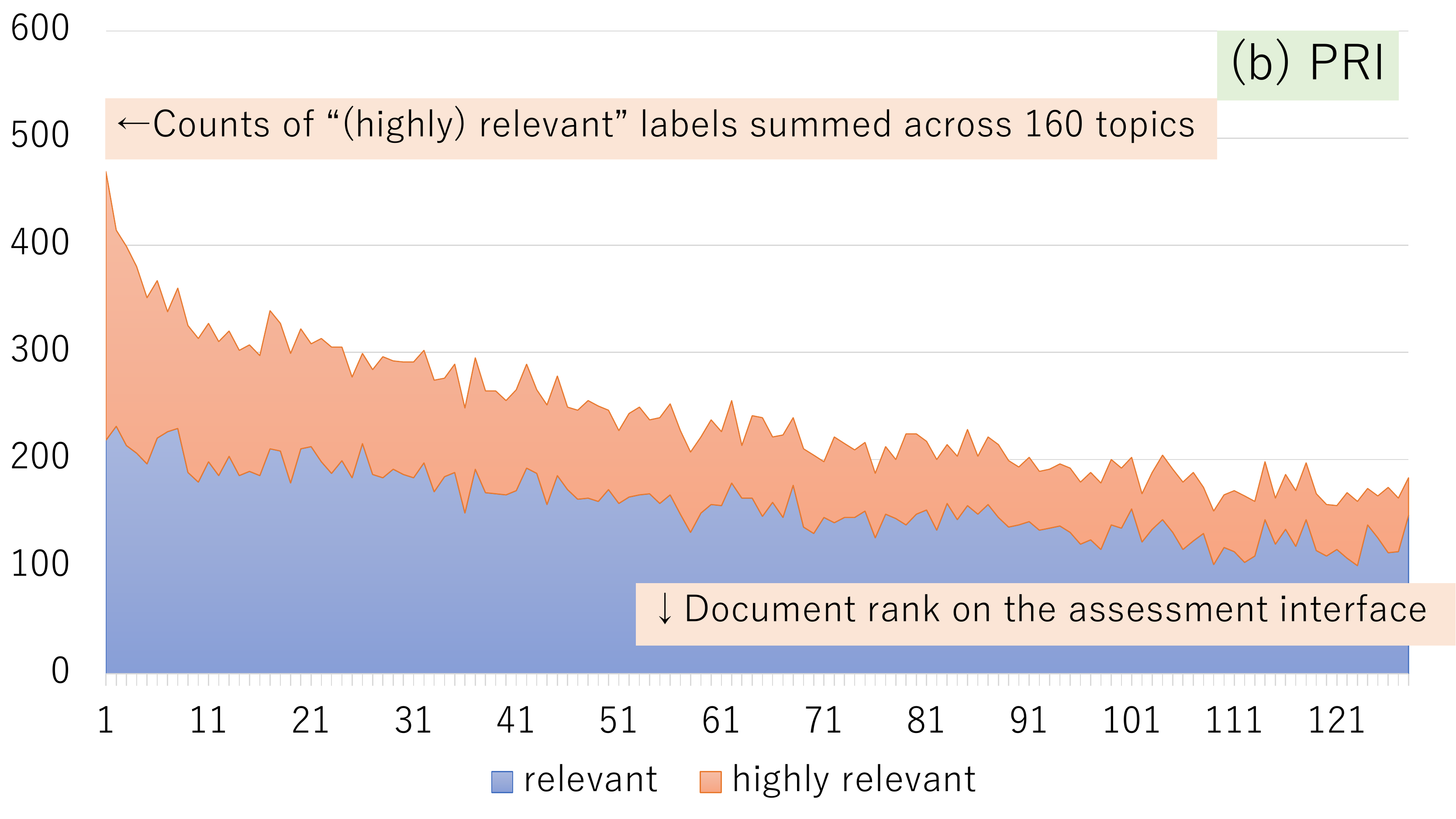}
\caption{Document presentation order vs. counts of relevance labels (relevant, highly relevant) 
based on the main experiment with the 160 topics.}\label{f:order-sum}

\end{center}
\end{figure}

This section discusses why different 
PRI-based qrels files produce relatively similar rankings.
To examine this phenomenon closely,
Figure~\ref{f:order-sum} visualises 
the relationship between the total number of \textbf{highly relevant} and \textbf{relevant} labels obtained
and the document presentation order as seen by the assessors
when WWW3E8 was constructed.
The $x$-axis represents the document ranks shown
on the PLY assessment interface.
As the \emph{minimum} pool size across the 160 topics was 128 (i.e., every topic had at least 128 pooled documents),
we count the assessors' labels (\textbf{highly relevant} or \textbf{relevant})
across the 160 topics for ranks 1-128.
As each topic was judged by four assessors 
for each document ordering strategy,
the maximum possible value for the $y$-axis 
is $4*160=640$: 
this would happen 
if, at a particular rank,
all four assessors
gave a \textbf{highly relevant} or \textbf{relevant} label
 for all 160 topics.
 Note that we are using the full topic set for this analysis
 as it does not involve computation of evaluation measures.

It is clear from Figure~\ref{f:order-sum} that,
while the trend is not very clear under the RND condition,
we obtain more and more \textbf{highly relevant} and \textbf{relevant} labels 
at higher ranks under the PRI condition.
There are two possible (mutually nonexclusive) explanations for this phenomenon:
(I)~the pseudorelevance as computed by \texttt{NTCIRPOOL}\footnote{
\url{http://research.nii.ac.jp/ntcir/tools/ntcirpool-en.html}
}
(based on the number of 
runs that returned that document and the ranks of that document) 
is
accurate to some degree, and often manages to present
truly relevant documents before nonrelevant ones;
(II)~under the PRI condition, the
assessors tend to \emph{overrate} the documents that they encounter early.
Recall that, for each topic,
each assessor receives either a RND pool or a PRI pool at random;
they are not even aware that there are two kinds of document ordering strategies.
The sharp contrast shown in  Figure~\ref{f:order-sum} despite this blind nature of the experiment
suggests that
the PRI strategy tends to prioritise documents that immediately \emph{look} relevant,
and that the assessors actually label them as so.
Again,
note that this does not rule out Explanation~(I):
the documents that \emph{look} relevant may often be truly relevant.

In summary,
\emph{while PRI-based qrels files tend to generate system ranking that are slightly more similar to each other
than RND-based qrels files do,
this difference is probably of no practical significance.
On the other hand, 
a PRI-based system ranking and a RND-based system ranking
can be quite different relative to the above within-PRI and within-RND ranking comparisons.
The PRI strategy tends to make the assessor favour ``popular'' documents,
i.e., those returned at high ranks by many systems.
}

\section{RQ4: Robustness to New Systems}\label{s:newsystems}

We now know that RND-based and PRI-based labels substantially differ from each other,
and that the system ranking similarities under the PRI condition 
tend to be slightly higher than those under the RND condition.
However, a more practically important question is
\textbf{RQ4}: which strategy is more robust to the evaluation of systems 
that did not contribute to the pool?
It is known that relevance assessments of test collections (especially those based on a small pool depth)
are \emph{incomplete},
and that new systems tend to be underrated if evaluated with such collections~\cite{sakai08cikm,sakai12airs,sakai14promise,voorhees02,zobel98},
because the new systems may return relevant documents 
that are outside the pools.
While researchers should be aware of this,
we still would not want test collections to fail catastrophically when evaluating new systems.

\begin{table}[t]
\centering

\caption{Number of topicdocs in leave-one-team-out qrels files.
The original qrels size covering the 147 topics is 29,522.
}\label{t:loto-qrels}
%\begin{scriptsize}
\begin{tabular}{c|c|r|c}
\hline
team left out &\#runs 	&unique 		&\#topicdocs\\
			&		&contributions	&in LOTO qrels\\
\hline
Group~1 		&3		&1,612		&27,910\\
Group~2		&5		&1,347		&28,175\\
Group~3		&5		&1,351		&28,171\\
Group~4		&5		&5,857		&23,665\\
Group~5		&5		&3,651		&25,871\\
Group~6		&5		&3,799		&25,723\\
Group~7		&5		&858		&28,664\\
Group~8		&1		&124		&29,398\\
Group~9		&3		&468		&29,054\\
\hline
\end{tabular}
%\end{scriptsize}
%\end{table}

%\begin{table}[t]
%\centering

%\vspace*{5mm}\caption{Mean system ranking $\tau$ over nine leave-one-team out experiments.
%For example, the RND1 column compares the original RND1 qrels
%with nine leave-one-team-out versions of the qrels.
%For each evaluation measure, a paired Tukey HSD test at the 5\% significance level was conducted. 
%$V_{E2}$ denotes the two-way ANOVA residual variance for computing effect sizes.
%}\label{t:newsystems}
%\begin{tabular}{c||c|c|c|c||c|c|c|c|c}
%\hline
%		&RND1	&RND2	&RND3	&RND4	&PRI1	&PRI2	&PRI3	&PRI4	&$V_{E2}$\\
%\hline
%nDCG	&0.914 	&0.893 	&0.901 	&0.910 	&0.956 	&0.959 	&0.957 	&0.965	&0.000876\\
%Q		&0.898 	&0.876 	&0.884 	&0.902 	&0.940 	&0.947 	&0.943 	&0.959	&0.001340\\
%nERR	&0.927 	&0.905 	&0.912 	&0.922 	&0.969 	&0.973 	&0.972 	&0.975	&0.000600\\
%iRBU	&0.923 	&0.910 	&0.907 	&0.901 	&0.965 	&0.966 	&0.957 	&0.966	&0.000859\\
%\hline
%\end{tabular}

\vspace*{5mm}\caption{Mean system ranking $\tau$ over nine leave-one-team out experiments.
For example, the RND1 column compares the original RND1 qrels
with nine leave-one-team-out versions of the qrels.
For each evaluation measure, a paired Tukey HSD test at the 5\% significance level was conducted. 
$V_{\mathrm{E2}}$ denotes the two-way ANOVA residual variance for computing effect sizes.
}\label{t:newsystems}
\begin{tabular}{c||c|c|c|c||c|c|c|c|c}
\hline
		&RND1	&RND2	&RND3	&RND4	&PRI1	&PRI2	&PRI3	&PRI4	&$V_{\mathrm{E2}}$\\
\hline
nDCG	&0.946 	&0.943 	&0.948 	&0.948 	&0.956 	&0.959 	&0.957 	&0.965 	&0.000173\\
Q		&0.937 	&0.932 	&0.935 	&0.933 	&0.940 	&0.947 	&0.943 	&0.959 	&0.000181\\
nERR	&0.956 	&0.956 	&0.957 	&0.956 	&0.969 	&0.973 	&0.972 	&0.975 	&0.000253\\ 
iRBU	&0.940 	&0.950 	&0.943 	&0.941 	&0.966 	&0.966 	&0.957 	&0.966 	&0.000317\\ 
\hline
\end{tabular}

\caption{All statistically significantly different pairs of qrels versions in terms of robustness to new systems (mean $\tau$
over $n=9$ leave-one-team-out experiments),
based on a paired Tukey HSD test at the 5\% significance level.
Effect sizes are based on the two-way ANOVA residual variances shown in Table~\ref{t:newsystems}.
None of the differences were statistically significant for nERR and iRBU.
}\label{t:newsystems-sig}
%\begin{small}
\begin{tabular}{c|l|c||c|l|c}
\hline
qrels pairs 		&$p$-value	&effect			&qrels pairs 		&$p$-value	&effect\\
				&			&size			&				&			&size\\
\hline
\multicolumn{3}{c||}
{(a) nDCG}&
\multicolumn{3}{c}
{(b) Q}\\
\hline
PRI4-RND2	&0.012		&1.73	&PRI4-RND2	&0.0022	&1.99\\
			&			&		&PRI4-RND4	&0.0040	&1.90\\
			&			&		&PRI4-RND3	&0.011	&1.74\\
			&			&		&PRI4-RND1	&0.020	&1.64\\
\hline
\end{tabular}

\end{table}

The robustness to new systems can be quantified using \emph{Leave-One-Team-Out} (LOTO) tests~\cite{sakai12airs,sakai14promise,voorhees02}.
That is, for each of the eight versions of qrels and for each team ($G$) that participated in the NTCIR-15 WWW-3 task~\cite{sakai20www3},
we remove $G$'s \emph{unique contributions} from the original qrels to form  a ``leave-out-$G$'' qrels file.
Here, a unique contribution is a topicdoc that was originally contributed to the pool by team $G$ and by no other team.
The WWW-3 task received runs from nine teams,
and therefore we created
$8*9=72$ LOTO qrels files
from the full qrels files covering the aforementioned 147 topics.
Table~\ref{t:loto-qrels} shows the relevant statistics of our LOTO experiments.
For example, by removing the 1,612 unique contributions of Group~1 from the full qrels file
that contained 29,522 topicdocs, 
we create a ``leave-out-Group-1'' qrels,
with which we can simulate a situation where ``new'' runs from Group~1
are evaluated using an existing test collection.
We then compare the system ranking based on the original qrels
with the new ranking based on each LOTO qrels in terms of Kendall's $\tau$.
If the $\tau$ is low, that means that the LOTO qrels substantially underrate the ``new'' runs,
which by extension suggests that the original qrels file is also not robust to real new runs
that did not contribute to the pools.

Table~\ref{t:newsystems} shows, for each of the eight qrels files (PRI1 through RND4) and for each evaluation measure,
the mean $\tau$ scores averaged over the nine LOTO trials.
Table~\ref{t:newsystems-sig} shows the accompanying results 
of the paired Tukey HSD tests.
%They can be summarised as follows.
%\begin{itemize}
%\item None of the differences \emph{within} each document ordering strategy are statistically significant.
%\item All statistically significant differences are cases where a PRI-based qrels file outperforms a RND-based qrels file.
%The largest effect size observed is over 2.0 for every evaluation measure,
%e.g., 2.86 for nERR and 2.21 for iRBU as shown in Table~\ref{t:newsystems-sig}(c) and~(d).
%\end{itemize}
%Hence the answer to \textbf{RQ4} is clear: \emph{the PRI strategy
%substantially outperforms the RND strategy in terms of robustness to new systems}.
%The result suggests that the PRI strategy often helps the assessors identify 
%relevant documents \emph{that affect the evaluation of many systems,
%regardless of 
%whether the systems contributed to the pools or not.}
%Put another way, because
%the RND strategy ignores
% the ``popularity'' of documents,
%it is liable 
%to miss relevant documents 
%that are useful for evaluating many systems fairly.
The following observations can be made from these results.
\begin{itemize}
\item For each evaluation measure,
the mean $\tau$'s for the PRI-based qrels files are slightly
higher than those for the RND-based qrels files, without exceptions.
\item Some of the mean differences are statistically significant at the 5\% significance level:
in particular, with Q-measure, the PRI4 qrels files is statistically significantly more robust to new systems 
than any of the RND-based qrels files.
More importantly, the effect sizes (i.e., standardised mean differences) are over 1.5
in each case, which suggests that the differences may be substantial.
\end{itemize}
Although not shown in Table~\ref{t:newsystems-sig},
we also observed ``almost statistically significant'' differences with iRBU as well:
PRI2 vs. RND1 ($p=0.053, \textit{ES}_{\mathrm{E2}}=1.47$),
PRI1 vs. RND1 ($p=0.059, \textit{ES}_{\mathrm{E2}}=1.45$), and
PRI4 vs. RND1 ($p=0.055, \textit{ES}_{\mathrm{E2}}=1.47$).
Hence, PRI4 is not necessarily the only robust PRI-based qrels file.
%(See also our additional ``depth-5'' results in Section~\ref{ss:additional-rq4}.)

\begin{figure}[t]
\begin{center}

\includegraphics[width=\textwidth]{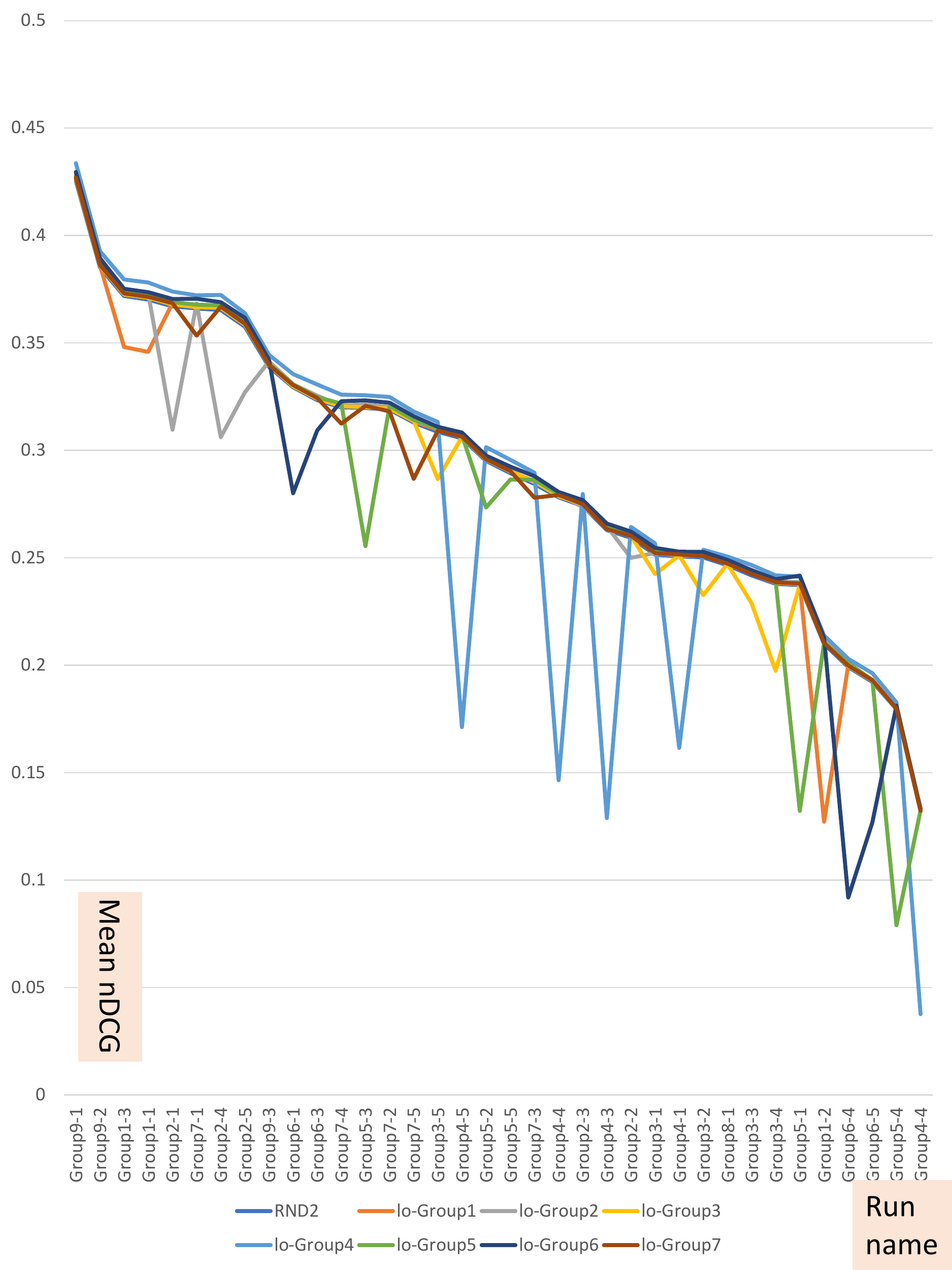}
\caption{Effect of leaving one team out from the RND2 qrels (Mean nDCG over 147 topics).}\label{f:RND2loto2}

\end{center}
\end{figure}

\begin{figure}[t]
\begin{center}

\includegraphics[width=\textwidth]{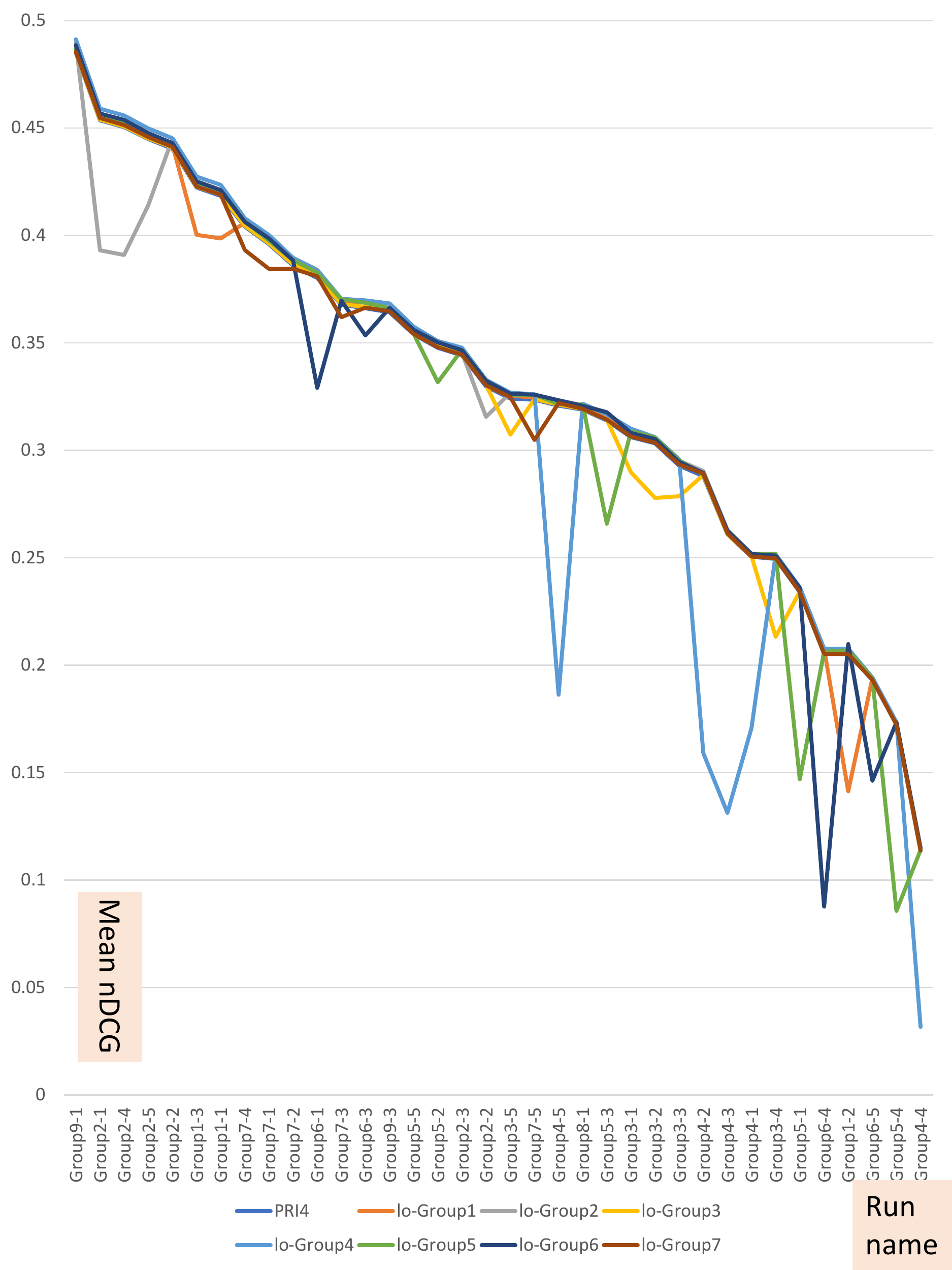}
\caption{Effect of leaving one team out from the PRI4 qrels (Mean nDCG over 147 topics).}\label{f:PRI4loto2}

\end{center}
\end{figure}

To examine the above result more closely,
Figures~\ref{f:RND2loto2} and~\ref{f:PRI4loto2} visualise the LOTO results with RND2 and PRI4 for nDCG,
whose mean $\tau$'s are the lowest and the highest among the eight versions of qrels
 %(0.893 
 (0.943
  and 0.965 as shown in Table~\ref{t:newsystems}, respectively).
The $y$-axis represents the mean nDCG scores,
while the $x$-axis represents the runs from all nine groups sorted according to the full qrels file over the 147 topics.
For example, ``Group1-1'' means Run~1 from Group~1.
The results of leaving out Groups 8 and~9
are omitted as their unique contributions are small (See Table~\ref{t:loto-qrels})
and therefore the curves are very similar to the one for the full qrels.
Runs that are heavily underrated by a LOTO qrels file can be identified as a ``V'' in the curves.\footnote{
This visualisation approach for LOTO tests was used earlier by \citet{sakai12airs} for the purpose of 
evaluating the robustness of diversified search evaluation measures.
}
For example, in Figure~\ref{f:PRI4loto2},
it is easy to observe from the ``lo-Group4'' curve 
that if Group~4 is left out,
this group's runs (e.g., Group4-5) are heavily underrated.
Leaving out this particular group disrupts the ranking this much because
this group had as many as 
5,857
unique contributions to the pools (See Table~\ref{t:loto-qrels}).
If we compare Figures~\ref{f:RND2loto2} and~\ref{f:PRI4loto2},
%it can be observed that:
%\begin{itemize}
%\item[(i)] Compared to PRI4, 
%RND2
%gives similar scores to all runs, suggesting that 
%PRI is indeed biased towards ``popular'' documents;
%\item[(ii)] For both RND2 and PRI4,
%the top half of the runs do not suffer much even when they are left out from the pools;
%it is the bottom half of the runs that are heavily underrated when treated as new runs.
%Moreover, the LOTO qrels files with RND2 suffer from this more often (i.e., there are more large ``V'''s).
%\end{itemize}
it can be observed that
large V's (i.e., substantial underestimation of new systems) 
are more or less evenly distributed across the $x$-axis for RND2,
while we only see smaller V's in the top half of the runs for PRI4.
That is, PRI4 is quite
robust to the evaluation of new systems that are actually effective.
As we have discussed in \citet{sakai22tois},
this is probably because the PRI strategy 
indeed helps us collect
``popular'' relevant documents,
i.e., relevant documents returned by many systems at high ranks,
which are also likely to be retrieved by new systems.

%Based on the above LOTO results that show the robustness of the PRI-based pools,
%if researchers need to evaluate their current IR systems using legacy IR test collections,
%we recommend the use of those constructed using the PRI approach
%unless they have a good reason to believe that
%their systems retrieve relevant documents that are vastly different
%from the pooled documents.
%Put another way,
%if the researchers believe that
%their new systems return search results 
%that are reasonably similar to existing systems,
%then a PRI-based test collection is recommended;
%otherwise the new systems may be heavily underrated.

In summary,
\emph{PRI-based qrels files tend to be slightly more robust to new systems than RND-based ones.
This is probably because the PRI strategy tends to help us identify ``popular'' relevant documents.
The ``popular'' relevant documents affect the evaluation of many systems,
including systems that did not contribute to the pools.
}

\clearpage

\section{Additional Experiments: Pool Depth and Run Quality}\label{s:additional}

This section reports on additional experiments that
utilise \emph{subsets} of the RND-based and PRI-based qrels files 
to re-examine \textbf{RQ3} (system ranking agreement discussed in Section~\ref{s:system-ranking})
and \textbf{RQ4} (robustness to new systems discussed in Section~\ref{s:newsystems}).
The additional experiments were conducted to address two specific questions
from a TOIS reviewer: 
(I)~\textit{What happens to the difference between RND and PRI 
if the pool depth is smaller?}
(II)~\textit{What happens to the difference between RND and PRI 
if the submitted runs are less effective?}
We did not incorporate these additional results into Sections~\ref{s:system-ranking} and~\ref{s:newsystems}
because
(a)~the experimental settings of the additional experiments are somewhat artificial,
which prevents us from making strong claims;
(b)~the results of these additional experiments are similar to our main experiments for \textbf{RQ3} and \textbf{RQ4}
and therefore do not affect our main conclusions; and
(c)~we wanted to maintain the conciseness of Sections~\ref{s:system-ranking} and~\ref{s:newsystems}
which address our original research questions.

\subsection{Method}\label{ss:additional-method}

%32375 ntcir15www2+3RND4.qrels
% 11989 137topics.RND1rr1-5.qrels
% 15089 137topics.RND1rr11-15.qrels

Recall that our main experiments for \textbf{RQ3} (system ranking agreement)
and \textbf{RQ4} (robustness to new systems) were based on depth-15 pools,
with 29,522 topicdocs covering 147 topics
 (See Section~\ref{ss:system-tau}).
We address Reviewer Question~(I) (effect of reducing the pool depth) as follows.
\begin{enumerate}
\item Form a list of topicdocs 
by extracting all documents from each of the 37 submitted run files,
where the document ranks are between 1 to 5.
\item Obtain a subset of each RND-based and PRI-based qrels file,
by using the above topicdoc list as a filter.
\end{enumerate}
Thus, although our relevance labels are still from the depth-15 based experiments,
the labelled documents we utilise in this additional experiments 
are those that qualify even if the pool depth was 5.
Note that this only \emph{approximates} the depth-5 situation
because (a)~the assessors processed the \emph{depth-15} pools in the presented order;
and (b)~for PRI-based pools, the pseudorelevance scores
based on depth-15 pools and those based on depth-5 pools generally differ,
since the number of runs that contain a document at or above rank 15
is generally not the same as the number of runs that contain that document at or above rank 5.
Nevertheless, the above setting is probably a reasonable approximation.
We shall refer to these filtered qrels files as \textbf{rr1-5} versions,
where \textbf{rr} stands for ``ranks in the runs.''

We addressed Reviewer Question~(II) (effect of poorly performing runs) in a similar way:
instead of extracting documents ranked between~1 and~5 in Step~(1) discussed above,
we extracted documents ranked between~11 and~15 (i.e., documents presumed worst 
in the depth-15 pool files).
%\footnote{
%This method was suggested by the reviewer.
%}
This \emph{approximates} a situation where the effectiveness of the runs are relatively poor
and different runs return different documents,
which should disrupt the majority-voting approach of PRI.
We shall refer to these filtered qrels files as \textbf{rr11-15} versions.
While this experiment is also a rough simulation of a real situation,
note that by comparing the outcomes of our \textbf{rr11-15} and \textbf{rr1-5} experiments,
we can discuss the effect of the run quality under the same condition,
where each run contributes exactly 5 documents to the pool.

In the additional experiments reported below, evaluation measure scores are averaged over 137 topics (a subset of the aforementioned 147 topics)
to avoid topics with zero relevant documents in the  \textbf{rr1-5} and \textbf{rr11-15} versions of the qrels files
as well as in their LOTO versions.
The 23 topics excluded are:
0012, 0024, 0026, 0027, 0031, 0044, 0047, 0058, 0060, 0063, 0076, 0132, 0143, 0144, 0147, 0150, 0152, 0153, 0165, 0169, 0174, 0175, 0179.\footnote{
This list is slightly different from the one from our original TOIS paper due to the aforementioned noise in the qrels files,
although that paper also used 137 topics for the additional experiments.
}
% for the following reasons.
%(1)~As our \textbf{rr1-5} and \textbf{rr11-15} versions of the qrels files
%are subsets of the original ones, for some topics
%we ended up with no relevant documents.
%The union of such topics across all 16 (i.e., 8 \textbf{rr1-5} and 8 \textbf{rr11-15}) qrels files
%amounts to 17 topics (0012,0024,0026,0027,0044,0060,0063,0132,0144,0147,0152,0153,0163,0169,0174,0177,0179).
%(2)~Furthermore, in our LOTO experiments based on the above reduced versions of the qrels files,
%we lost 6 topics in addition (0014,0047,0058,0143,0150,0175).
%Hence, in order to evaluate the runs using a common topic set for every experimental condition,
%we average the evaluation measures over $160-17-6=137$ topics.
The \textbf{rr1-5} qrel files each contain 11,989 topicdocs covering the 137 topics,
whereas, the \textbf{rr11-15} qrels files each contain 15,089 topicdocs.
The latter number is larger because the documents ranked between 11 and 15 by each run
are indeed less similar than those ranked between 1 and 5 by each run.
That is, the tendency with the \textbf{rr11-15} setting is indeed that
different runs contribute different documents.

\begin{table}[t]
\centering
\begin{small}

\caption{System ranking agreement as measured by Kendall's $\tau$ between system ranking pairs ($n=36$ runs)
according to the four official measures of the NTCIR-15 WWW-3 task, based on the \textbf{rr1-5} versions
of the qrels (averaged over 137 topics).
}\label{t:inter-qrels-tau-rr1-5}
\begin{tabular}{c|c|c|c|c|c|c|c}
\hline
(a) nDCG &RND2	&RND3	&RND4	&PRI1	&PRI2	&PRI3	&PRI4\\
\hline
RND1	&0.913	&0.902	&0.921	&0.930	&0.949	&0.898	&0.900\\
RND2	&-		&0.903	&0.897	&0.862	&0.900	&0.856	&0.825\\
RND3	&-		&-		&0.937	&0.895	&0.895	&0.857	&0.852\\
RND4	&-		&-		&-		&0.927	&0.921	&0.889	&0.878\\
PRI1	&-		&-		&-		&-		&0.930	&0.937	&0.925\\
PRI2	&-		&-		&-		&-		&-		&0.911	&0.906\\
PRI3	&-		&-		&-		&-		&-		&-		&0.919\\
\hline
(b) Q 	&RND2	&RND3	&RND4	&PRI1	&PRI2	&PRI3	&PRI4\\
\hline
RND1	&0.917	&0.917	&0.908	&0.937	&0.949	&0.927	&0.895\\
RND2	&-		&0.930	&0.908	&0.879	&0.905	&0.870	&0.851\\
RND3	&-		&-		&0.914	&0.886	&0.917	&0.876	&0.851\\
RND4	&-		&-		&-		&0.889	&0.914	&0.873	&0.886\\
PRI1	&-		&-		&-		&-		&0.943	&0.933	&0.914\\
PRI2	&-		&-		&-		&-		&-		&0.933	&0.902\\
PRI3	&-		&-		&-		&-		&-		&-		&0.911\\
\hline
(c) nERR &RND2	&RND3	&RND4	&PRI1	&PRI2	&PRI3	&PRI4\\
\hline
RND1	&0.825	&0.851	&0.851	&0.860	&0.851	&0.838	&0.824\\
RND2	&-		&0.860	&0.810	&0.781	&0.771	&0.746	&0.757\\
RND3	&-		&-		&0.867	&0.806	&0.810	&0.765	&0.786\\
RND4	&-		&-		&-		&0.870	&0.867	&0.835	&0.849\\
PRI1	&-		&-		&-		&-		&0.876	&0.895	&0.903\\
PRI2	&-		&-		&-		&-		&-		&0.898	&0.890\\
PRI3	&-		&-		&-		&-		&-		&-		&0.897\\								
\hline
(d) iRBU &RND2	&RND3	&RND4	&PRI1	&PRI2	&PRI3	&PRI4\\
\hline
RND1	&0.875	&0.837	&0.884	&0.890	&0.881	&0.903	&0.914\\
RND2	&-		&0.886	&0.914	&0.832	&0.854	&0.844	&0.856\\
RND3	&-		&-		&0.895	&0.832	&0.829	&0.832	&0.843\\
RND4	&-		&-		&-		&0.848	&0.832	&0.873	&0.878\\
PRI1	&-		&-		&-		&-		&0.870	&0.911	&0.906\\
PRI2	&-		&-		&-		&-		&-		&0.908	&0.881\\
PRI3	&-		&-		&-		&-		&-		&-		&0.916\\
\hline
\end{tabular}

%\caption{Comparison of mean system ranking $\tau$'s based on the $\tau$'s shown in Table~\ref{t:inter-qrels-tau-rr1-5}.
%Results of
%Tukey HSD tests for unpaired data with sample sizes 6, 6, 16 are shown.
%The effect sizes are standardised mean differences based on the 
%one-way ANOVA
%residual variance $V_{E1}$~\cite{sakai18book}.
%}\label{t:tau-significance-rr1-5}
%\begin{tabular}{c|c|c|c|c|c|c|l}
%\hline
%Measure	&\multicolumn{3}{|c|}{Mean $\tau$ (sample size)}	&Residual	&\multicolumn{3}{|c}{$p$-value (effect size)}\\
%		&RND-RND		&PRI-PRI		&RND-PRI		&variance	&RND-RND vs	&PRI-PRI vs	&PRI-PRI vs\\
%		&($n_{1}=6$)	&($n_{1}=6$)	&($n_{1}=16$)	&$V_{E1}$	&RND-PRI		&RND-PRI		&RND-RND\\
%\hline
%nDCG	&0.847		&0.916		&0.471		&0.000659	&$\approx 0$ (14.7)	&$\approx 0$ (17.3)	&0.000269 (2.68)\\
%Q		&0.847		&0.923		&0.503		&0.000404	&$\approx 0$ (17.1)	&$\approx 0$ (20.9)	&0.000002 (3.78)\\
%nERR	&0.659		&0.889		&0.311		&0.002012	&$\approx 0$ (7.77)	&$\approx 0$ (12.9)	&$\approx 0$ (5.12)\\
%iRBU	&0.755		&0.891		&0.478		&0.002833	&$\approx 0$ (5.20)	&$\approx 0$ (7.76)	&0.000457 (2.56)\\
%\hline
%\end{tabular}

\caption{Comparison of mean system ranking $\tau$'s based on the $\tau$'s shown in Table~\ref{t:inter-qrels-tau-rr1-5}
with
Tukey HSD tests for unpaired data.
Statistical significance at $\alpha=0.05$ is indicated in \textbf{bold}.
The effect sizes are standardised mean differences based on the 
one-way ANOVA
residual variance $V_{\mathrm{E1}}$~\cite{sakai18book}.
}\label{t:tau-significance-rr1-5}
\begin{tabular}{c|c|c|c|c|c|c|l}
\hline
Measure	&\multicolumn{3}{|c|}{Mean $\tau$ (sample size)}	&Residual	&\multicolumn{3}{|c}{$p$-value (effect size)}\\
		&RND-RND		&PRI-PRI		&RND-PRI		&variance	&RND-RND vs	&PRI-PRI vs	&PRI-PRI vs\\
		&($n=6$)	&($n=6$)	&($n=16$)	&$V_{\mathrm{E1}}$	&RND-PRI		&RND-PRI		&RND-RND\\
\hline
nDCG	&0.912		&0.921		&0.890		&0.000734	&$p=0.21$			&$p=0.055$			&$p=0.83$\\
		&			&			&			&			&$\textit{ES}=0.83$	&$\textit{ES}=1.17$	&$\textit{ES}=0.34$\\
Q		&0.916		&0.923		&0.894		&0.000561	&$p=0.16$			&$\mathbf{p=0.047}$			&$p=0.87$\\
		&			&			&			&			&$\textit{ES}=0.91$	&$\textit{ES}=1.21$	&$\textit{ES}=0.30$\\
nERR	&0.844		&0.893		&0.814		&0.00114		&$p=0.17$			&$\mathbf{p<0.0002}$			&$\mathbf{p=0.048}$\\
		&			&			&			&			&$\textit{ES}=0.90$	&$\textit{ES}=2.34$	&$\textit{ES}=1.46$\\
iRBU	&0.882		&0.899		&0.859		&0.000657	&$p=0.17$			&$\mathbf{p=0.0089}$			&$p=0.50$\\
		&			&			&			&			&$\textit{ES}=0.90$	&$\textit{ES}=1.55$	&$\textit{ES}=0.66$\\
\hline
\end{tabular}

\end{small}
\end{table}

\begin{table}[t]
\centering

\begin{small}

\caption{System ranking agreement as measured by Kendall's $\tau$ between system ranking pairs ($n=36$ runs)
according to the four official measures of the NTCIR-15 WWW-3 task, based on the \textbf{rr11-15} versions
of the qrels (averaged over 137 topics).
}\label{t:inter-qrels-tau-rr11-15}
\begin{tabular}{c|c|c|c|c|c|c|c}
\hline
(a) nDCG &RND2	&RND3	&RND4	&PRI1	&PRI2	&PRI3	&PRI4\\
\hline
RND1	&0.911	&0.917	&0.908	&0.921	&0.921	&0.911	&0.876\\
RND2	&-		&0.911	&0.857	&0.883	&0.921	&0.879	&0.857\\
RND3	&-		&-		&0.921	&0.883	&0.908	&0.873	&0.876\\
RND4	&-		&-		&-		&0.892	&0.886	&0.870	&0.892\\
PRI1	&-		&-		&-		&-		&0.930	&0.946	&0.930\\
PRI2	&-		&-		&-		&-		&-		&0.921	&0.924\\
PRI3	&-		&-		&-		&-		&-		&-		&0.933\\
\hline
(b) Q 	&RND2	&RND3	&RND4	&PRI1	&PRI2	&PRI3	&PRI4\\
\hline
RND1	&0.921	&0.951	&0.933	&0.924	&0.946	&0.892	&0.895\\
RND2	&-		&0.913	&0.898	&0.883	&0.924	&0.863	&0.879\\
RND3	&-		&-		&0.954	&0.925	&0.951	&0.894	&0.894\\
RND4	&-		&-		&-		&0.914	&0.937	&0.889	&0.892\\
PRI1	&-		&-		&-		&-		&0.933	&0.917	&0.902\\
PRI2	&-		&-		&-		&-		&-		&0.908	&0.917\\
PRI3	&-		&-		&-		&-		&-		&-		&0.933\\
\hline
(c) nERR &RND2	&RND3	&RND4	&PRI1	&PRI2	&PRI3	&PRI4\\
\hline
RND1	&0.927	&0.919	&0.889	&0.943	&0.892	&0.878	&0.870\\
RND2	&-		&0.916	&0.886	&0.908	&0.883	&0.856	&0.860\\
RND3	&-		&-		&0.906	&0.887	&0.887	&0.832	&0.862\\
RND4	&-		&-		&-		&0.889	&0.902	&0.852	&0.898\\
PRI1	&-		&-		&-		&-		&0.917	&0.910	&0.902\\
PRI2	&-		&-		&-		&-		&-		&0.887	&0.902\\
PRI3	&-		&-		&-		&-		&-		&-		&0.916\\
\hline
(d) iRBU &RND2	&RND3	&RND4	&PRI1	&PRI2	&PRI3	&PRI4\\
\hline
RND1	&0.886	&0.892	&0.916	&0.898	&0.883	&0.902	&0.868\\
RND2	&-		&0.937	&0.929	&0.917	&0.902	&0.895	&0.897\\
RND3	&-		&-		&0.910	&0.905	&0.863	&0.889	&0.890\\
RND4	&-		&-		&-		&0.929	&0.922	&0.887	&0.937\\
PRI1	&-		&-		&-		&-		&0.895	&0.895	&0.929\\
PRI2	&-		&-		&-		&-		&-		&0.892	&0.916\\
PRI3	&-		&-		&-		&-		&-		&-		&0.884\\
\hline
\end{tabular}

%\caption{Comparison of mean system ranking $\tau$'s based on the $\tau$'s shown in Table~\ref{t:inter-qrels-tau-rr11-15}.
%Results of
%Tukey HSD tests for unpaired data with sample sizes 6, 6, 16 are shown.
%The effect sizes are standardised mean differences based on the 
%one-way ANOVA
%residual variance $V_{E1}$~\cite{sakai18book}.
%}\label{t:tau-significance-rr11-15}
%\begin{tabular}{c|c|c|c|c|c|c|l}
%\hline
%Measure	&\multicolumn{3}{|c|}{Mean $\tau$ (sample size)}	&Residual	&\multicolumn{3}{|c}{$p$-value (effect size)}\\
%		&RND-RND		&PRI-PRI		&RND-PRI		&variance	&RND-RND vs	&PRI-PRI vs	&PRI-PRI vs\\
%		&($n_{1}=6$)	&($n_{1}=6$)	&($n_{1}=16$)	&$V_{E1}$	&RND-PRI		&RND-PRI		&RND-RND\\
%\hline
%nDCG	&0.878		&0.926		&0.593		&0.000355	&$\approx 0$ (15.1)	&$\approx 0$ (17.6)	&0.000488 (2.55)\\
%Q		&0.868		&0.913		&0.641		&0.000737	&$\approx 0$ (8.35)	&$\approx 0$ (10.0)	&0.0232 (1.64)\\
%nERR	&0.831		&0.904		&0.556		&0.001194	&$\approx 0$ (7.98)	&$\approx 0$ (10.1)	&0.00327 (2.11)\\
%iRBU	&0.818		&0.898		&0.604		&0.001801	&$\approx 0$ (5.02)	&$\approx 0$ (6.91)	&0.00858 (1.89)\\
%\hline
%\end{tabular}

\caption{Comparison of mean system ranking $\tau$'s based on the $\tau$'s shown in Table~\ref{t:inter-qrels-tau-rr11-15}.
with
Tukey HSD tests for unpaired data.
Statistical significance at $\alpha=0.05$ is indicated in \textbf{bold}.
The effect sizes are standardised mean differences based on the 
one-way ANOVA
residual variance $V_{\mathrm{E1}}$~\cite{sakai18book}.
}\label{t:tau-significance-rr11-15}
\begin{tabular}{c|c|c|c|c|c|c|l}
\hline
Measure	&\multicolumn{3}{|c|}{Mean $\tau$ (sample size)}	&Residual	&\multicolumn{3}{|c}{$p$-value (effect size)}\\
		&RND-RND		&PRI-PRI		&RND-PRI		&variance	&RND-RND vs	&PRI-PRI vs	&PRI-PRI vs\\
		&($n=6$)	&($n=6$)	&($n=16$)	&$V_{\mathrm{E1}}$	&RND-PRI		&RND-PRI		&RND-RND\\
\hline
nDCG	&0.904		&0.931		&0.891		&0.000369	&$p=0.32$ 			&$\mathbf{p<0.0006}$ 			&$p=0.062$\\
		&			&			&			&			&$\textit{ES}=0.71$	&$\textit{ES}=2.08$	&$\textit{ES}=1.38$\\
Q		&0.928		&0.918		&0.906		&0.000523	&$p=0.32$ 			&$p=0.13$ 			&$p=0.73$\\
		&			&			&			&			&$\textit{ES}=0.52$	&$\textit{ES}=0.96$	&$\textit{ES}=0.44$\\
nERR	&0.907		&0.906		&0.881		&0.000495	&$p=0.074$ 			&$p=0.056$ 			&$p=0.99$\\
		&			&			&			&			&$\textit{ES}=1.10$	&$\textit{ES}=1.17$	&$\textit{ES}=0.07$\\
iRBU	&0.912		&0.902		&0.899		&0.000383	&$p=0.95$ 			&$p=0.38$ 			&$p=0.66$\\
		&			&			&			&			&$\textit{ES}=0.14$	&$\textit{ES}=0.65$	&$\textit{ES}=0.50$\\
\hline
\end{tabular}

\end{small}

\end{table}

\subsection{Revisiting RQ3: System Ranking Agreement}

Table~\ref{t:inter-qrels-tau-rr1-5} shows the system ranking agreement results 
using the \textbf{rr1-5} (i.e., ``depth-5'') versions of the qrels,
and Table~\ref{t:tau-significance-rr1-5} shows the corresponding Tukey HSD test results 
for comparing the mean $\tau$ within RND, within PRI, and between RND and PRI.
These tables are arranged in the same way 
as Table~\ref{t:inter-qrels-tau} and Table~\ref{t:tau-significance}.
Similarly,
Table~\ref{t:inter-qrels-tau-rr11-15} shows the system ranking agreement results 
using the \textbf{rr11-15} (i.e., ``poor runs'') versions of the qrels,
and Table~\ref{t:tau-significance-rr11-15} shows the corresponding Tukey HSD test results.

In Section~\ref{s:system-ranking}, we concluded as follows.
\emph{While PRI-based qrels files tend to generate system ranking that are slightly more similar to each other
than RND-based qrels files do,
this difference is probably of no practical significance.
On the other hand, 
a PRI-based system ranking and a RND-based system ranking
can be quite different relative to the above within-PRI and within-RND ranking comparisons.
The PRI strategy tends to make the assessor favour ``popular'' documents,
i.e., those returned at high ranks by many systems.
}
Our \textbf{rr1-5} (shallow pools) and \textbf{rr11-15} (poor run quality) results
are generally in line with the main conclusions,
while the following two points may be worth noting.
\begin{itemize}
\item For the \textbf{rr1-5} nERR results (Table~\ref{t:tau-significance-rr1-5}),
the PRI-PRI system ranking comparisons statistically significant outperform the RND-RND ones ($p=0.048$),
despite the small sample sizes.
\item In the \textbf{rr11-15} results (Table~\ref{t:tau-significance-rr11-15}),
the mean $\tau$ values for RND-RND are \emph{not} smaller than the PRI-PRI ones for Q, nERR, and iRBU.
On the other hand, for nDCG, PRI-PRI almost statistically significantly outperforms RND-RND ($p=0.062, \textit{ES}=1.38$).
\end{itemize}

%\clearpage

\subsection{Revisiting RQ4: Robustness to New Systems}

We also reran the LOTO experiments for evaluating the robustness of qrels files to new systems (i.e., \textbf{RQ4}),
starting with the \textbf{rr1-5} (``depth-5'') and \textbf{rr11-15} (``poor runs'') qrels files
described in Section~\ref{ss:additional-method}.
Tables~\ref{t:loto-qrels-rr1-5}-\ref{t:newsystems-sig-r1-5}
show the results under the \textbf{rr1-5} setting;
they are arranged in exactly the same way 
as our main results (Tables~\ref{t:loto-qrels}-\ref{t:newsystems-sig}).
Similarly, Tables~\ref{t:loto-qrels-rr11-15}-\ref{t:newsystems-sig-r11-15}
show the results under the \textbf{rr11-15} setting.
%It can be observed that our new results are similar to the main results discussed in Section~\ref{s:newsystems}.
%That is, (1)~None of the differences \emph{within} each document ordering strategy are statistically significant; and
%(2)~All statistically significant differences are cases where a PRI-based qrels file outperforms 
%a RND-based qrels file. 

%In Section~\ref{s:newsystems}, we concluded that 
%\emph{the PRI strategy substantially outperforms the RND strategy in terms of robustness to new systems}, and that
%\emph{the PRI strategy often helps the assessors identify 
%relevant documents that affect the evaluation of many systems, regardless of 
%whether the systems contributed to the pools or not.}
%Our \textbf{rr1-5} and \textbf{rr11-15} results
%suggest that these conclusions may generalise to cases 
%where the pool depth is as small as five,
%and even if the runs are not very effective.
%Again, however, the reader is reminded that these 
%experiments only \emph{approximate} real situations with depth-5 pools and poorly performing runs.

In Section~\ref{s:newsystems}, we concluded as follows.
\emph{PRI-based qrels files tend to be slightly more robust to new systems than RND-based ones.
This is probably because the PRI strategy tends to help us identify ``popular'' relevant documents.
The ``popular'' relevant documents affect the evaluation of many systems,
including systems that did not contribute to the pools.}
Our \textbf{rr1-5} (shallow pools) and \textbf{rr11-15} (poor run quality) results
are generally in line with the main conclusions. More specifically:
\begin{itemize}
\item The \textbf{rr1-5} statistical significance results (Table~\ref{t:newsystems-sig-r1-5}) actually suggest more clearly 
than the depth-15 results (Table~\ref{t:newsystems-sig})
that PRI may be more robust to new systems than RND.
Every statistically significant difference 
suggests that a PRI qrels file is more robust than another qrels file.
We now have a few statistically significant differences even with nERR and iRBU.
\item The \textbf{rr11-15}  statistical significance results (Table~\ref{t:newsystems-sig-r11-15}) 
are less clear, in that there are 4 cases (2 cases in terms of Q and 2 cases in terms of iRBU)
where a RND qrels file statistically significantly outperforms a PRI file.
However, there are as many as 11 cases (9 in terms of nDCG, 1 in terms of Q, and 1 in terms of iRBU)
where a PRI qrels file statistically significantly outperforms a RND file.
\end{itemize}

\begin{table}[t]
\centering

\begin{small}

%\caption{Number of topicdocs in leave-one-team-out qrels files under the \textbf{rr1-5} setting (``depth-5'').
%The original qrels size covering the 137 topics is 11,989.
%}\label{t:loto-qrels-rr1-5}
%\begin{tabular}{c|c|r|c}
%\hline
%team left out &\#runs 	&unique 		&\#topicdocs\\
%			&		&contributions	&in LOTO qrels\\
%\hline
%Group~1 		&3		&714	&11,275\\
%Group~2		&5		&565	&11,424\\
%Group~3		&5		&1,019	&10,970\\
%Group~4		&5		&2,237	&9,752\\
%Group~5		&5		&1,503	&10,486\\
%Group~6		&5		&1,551	&10,438\\
%Group~7		&5		&570	&11,419\\
%Group~8		&1		&112	&11,877\\
%Group~9		&3		&456	&11,533\\
%\hline
%\end{tabular}

\caption{Number of topicdocs in leave-one-team-out qrels files under the \textbf{rr1-5} setting (``depth-5'').
The original qrels size covering the 137 topics is 
11,957.
}\label{t:loto-qrels-rr1-5}
\begin{tabular}{c|c|r|c}
\hline
team left out &\#runs 	&unique 		&\#topicdocs\\
			&		&contributions	&in LOTO qrels\\
\hline
Group~1 		&3		&714	&11,243\\
Group~2		&5		&549	&11,408\\
Group~3		&5		&1,011	&10.946\\
Group~4		&5		&2,236	&9,721\\
Group~5		&5		&1,496	&10,461\\
Group~6		&5		&1,551	&10,406\\
Group~7		&5		&564	&11,393\\
Group~8		&1		&113	&11,844\\
Group~9		&3		&460	&11,497\\
\hline
\end{tabular}

%\caption{Mean system ranking $\tau$ over nine leave-one-team out experiments under the \textbf{rr1-5} setting (``depth-5'').
%For example, the RND1 column compares the original RND1 qrels
%with nine leave-one-team-out versions of the qrels.
%For each evaluation measure, a paired Tukey HSD test at the 5\% significance level was conducted. 
%$V_{E2}$ denotes the two-way ANOVA residual variance for computing effect sizes.
%}\label{t:newsystems-r1-5}
%\begin{tabular}{c||c|c|c|c||c|c|c|c|c}
%\hline
%		&RND1	&RND2	&RND3	&RND4	&PRI1	&PRI2	&PRI3	&PRI4	&$V_{E2}$\\
%\hline
%nDCG	&0.895 	&0.878 	&0.874 	&0.890 	&0.925 	&0.932 	&0.927 	&0.942	&0.000885\\
%Q		&0.883 	&0.862 	&0.871 	&0.883 	&0.896 	&0.911 	&0.918 	&0.922	&0.001171\\
%nERR	&0.871 	&0.850 	&0.858 	&0.871 	&0.935 	&0.939 	&0.941 	&0.942	&0.001205\\
%iRBU	&0.892 	&0.891 	&0.877 	&0.878 	&0.941 	&0.934 	&0.944 	&0.940	&0.000894\\
%\hline
%\end{tabular}

\caption{Mean system ranking $\tau$ over nine leave-one-team out experiments under the \textbf{rr1-5} setting (``depth-5'').
For example, the RND1 column compares the original RND1 qrels
with nine leave-one-team-out versions of the qrels.
For each evaluation measure, a paired Tukey HSD test at the 5\% significance level was conducted. 
$V_{\mathrm{E2}}$ denotes the two-way ANOVA residual variance for computing effect sizes.
}\label{t:newsystems-r1-5}
\begin{tabular}{c||c|c|c|c||c|c|c|c|c}
\hline
		&RND1	&RND2	&RND3	&RND4	&PRI1	&PRI2	&PRI3	&PRI4	&$V_{\mathrm{E2}}$\\
\hline
nDCG	&0.923 	&0.921 	&0.925 	&0.927 	&0.925 	&0.932 	&0.934 	&0.943	&0.000108\\ 
Q		&0.908 	&0.897 	&0.908 	&0.909 	&0.898 	&0.914 	&0.920 	&0.926	&0.000110\\
nERR	&0.919 	&0.926 	&0.925 	&0.930 	&0.934 	&0.942 	&0.942 	&0.945	&0.000237\\
iRBU	&0.923 	&0.933 	&0.928 	&0.923 	&0.940 	&0.934 	&0.941 	&0.944	&0.000190\\ 
\hline
\end{tabular}

\caption{All statistically significantly different pairs of qrels versions (under the \textbf{rr1-5}) setting in terms of robustness to new systems (mean $\tau$
over $n=9$ leave-one-team-out experiments),
based on a paired Tukey HSD test at the 5\% significance level.
Effect sizes are based on the two-way ANOVA residual variances shown in Table~\ref{t:newsystems-r1-5}.
}\label{t:newsystems-sig-r1-5}
\begin{tabular}{c|l|c||c|l|c}
\hline
qrels pairs 		&$p$-value	&effect			&qrels pairs 		&$p$-value	&effect\\
				&			&size			&				&			&size\\
\hline
\multicolumn{3}{c||}
{(a) nDCG}&
\multicolumn{3}{c}
{(b) Q}\\
\hline
PRI4-RND2	&0.00078		&2.14	&PRI4-RND2 	&0.0000074	&2.75\\
PRI4-RND1 	&0.0034		&1.92	&PRI4-PRI1 	&0.000016	&2.66\\
PRI4-PRI1 	&0.015		&1.70	&PRI3-RND2 	&0.00052		&2.19\\
PRI4-RND3 	&0.017		&1.68	&PRI3-PRI1 	&0.0010		&2.10\\
PRI4-RND4 	&0.047		&1.50	&PRI4-RND3 	&0.011		&1.74\\
			&			&		&PRI4-RND1 	&0.011		&1.74\\
			&			&		&PRI4-RND4 	&0.022		&1.63\\
			&			&		&PRI2-RND2 	&0.025		&1.61\\
			&			&		&PRI2-PRI1 	&0.042		&1.51\\
\hline
\multicolumn{3}{c||}
{(c) nERR}&
\multicolumn{3}{c}
{(d) iRBU}\\
\hline
PRI4-RND1 &0.0174921	&1.67	&PRI4-RND1 &0.0363969	&1.54\\
			&			&		&PRI4-RND4 &0.0416764	&1.52\\
\hline
\end{tabular}

\end{small}

\end{table}

\begin{table}[t]
\centering

\begin{small}

%\caption{Number of topicdocs in leave-one-team-out qrels files under the \textbf{rr11-15} setting (``low run quality'').
%The original qrels size covering the 137 topics is 15,089.
%}\label{t:loto-qrels-rr11-15}
%\begin{tabular}{c|c|r|c}
%\hline
%team left out &\#runs 	&unique 		&\#topicdocs\\
%			&		&contributions	&in LOTO qrels\\
%\hline
%Group~1 		&3		&900	&14,189\\
%Group~2		&5		&1,046	&14,043\\
%Group~3		&5		&1,076	&14,013\\
%Group~4		&5		&2,529	&12,560\\
%Group~5		&5		&1,838	&13,251\\
%Group~6		&5		&1,928	&13,161\\
%Group~7		&5		&907	&14,182\\
%Group~8		&1		&120	&14,969\\
%Group~9		&3		&568	&14,521\\
%\hline
%\end{tabular}

\caption{Number of topicdocs in leave-one-team-out qrels files under the \textbf{rr11-15} setting (``low run quality'').
The original qrels size covering the 137 topics is 
15,057
}\label{t:loto-qrels-rr11-15}
\begin{tabular}{c|c|r|c}
\hline
team left out &\#runs 	&unique 		&\#topicdocs\\
			&		&contributions	&in LOTO qrels\\
\hline
Group~1 		&3		&902	&14,155\\
Group~2		&5		&1,030	&14,027\\
Group~3		&5		&1,064	&13,993\\
Group~4		&5		&2,525	&12,532\\
Group~5		&5		&1,845	&13,212\\
Group~6		&5		&1,924	&13,133\\
Group~7		&5		&911	&14,146\\
Group~8		&1		&121	&14,936\\
Group~9		&3		&560	&14,497\\
\hline
\end{tabular}

%\caption{Mean system ranking $\tau$ over nine leave-one-team out experiments under the \textbf{rr11-15} setting (``low run quality'').
%For example, the RND1 column compares the original RND1 qrels
%with nine leave-one-team-out versions of the qrels.
%For each evaluation measure, a paired Tukey HSD test at the 5\% significance level was conducted. 
%$V_{E2}$ denotes the two-way ANOVA residual variance for computing effect sizes.
%}\label{t:newsystems-r11-15}
%\begin{tabular}{c||c|c|c|c||c|c|c|c|c}
%\hline
%		&RND1	&RND2	&RND3	&RND4	&PRI1	&PRI2	&PRI3	&PRI4	&$V_{E2}$\\
%\hline
%nDCG	&0.949 	&0.944 	&0.955 	&0.966 	&0.978 	&0.977 	&0.975 	&0.980 	&0.000259\\
%Q		&0.950 	&0.947 	&0.955 	&0.959 	&0.970 	&0.965 	&0.957 	&0.974	&0.000323\\
%nERR	&0.938 	&0.940 	&0.949 	&0.946 	&0.970 	&0.974 	&0.980 	&0.972	&0.000235\\
%iRBU	&0.957 	&0.950 	&0.952 	&0.951 	&0.975 	&0.967 	&0.980 	&0.975	&0.000201\\
%\hline
%\end{tabular}

\caption{Mean system ranking $\tau$ over nine leave-one-team out experiments under the \textbf{rr11-15} setting (``low run quality'').
For example, the RND1 column compares the original RND1 qrels
with nine leave-one-team-out versions of the qrels.
For each evaluation measure, a paired Tukey HSD test at the 5\% significance level was conducted. 
$V_{\mathrm{E2}}$ denotes the two-way ANOVA residual variance for computing effect sizes.
}\label{t:newsystems-r11-15}
\begin{tabular}{c||c|c|c|c||c|c|c|c|c}
\hline
		&RND1	&RND2	&RND3	&RND4	&PRI1	&PRI2	&PRI3	&PRI4	&$V_{\mathrm{E2}}$\\
\hline
nDCG	&0.974 	&0.971 	&0.957 	&0.968 	&0.978 	&0.978 	&0.977 	&0.986	&0.0000384\\ 
Q		&0.970 	&0.970 	&0.977 	&0.974 	&0.971 	&0.972 	&0.958 	&0.975	&0.0000797\\ 
nERR	&0.971 	&0.971 	&0.967 	&0.966 	&0.973 	&0.972 	&0.978 	&0.972	&0.0000823\\ 
iRBU	&0.965 	&0.975 	&0.976 	&0.973 	&0.971 	&0.962 	&0.980 	&0.969	&0.0000567\\ 
\hline
\end{tabular}

\caption{All statistically significantly different pairs of qrels versions (under the \textbf{rr11-15}) setting in terms of robustness to new systems (mean $\tau$
over $n=9$ leave-one-team-out experiments),
based on a paired Tukey HSD test at the 5\% significance level.
Effect sizes are based on the two-way ANOVA residual variances shown in Table~\ref{t:newsystems-r11-15}.
}\label{t:newsystems-sig-r11-15}
\begin{tabular}{c|l|c||c|l|c}
\hline
qrels pairs 		&$p$-value	&effect			&qrels pairs 		&$p$-value	&effect\\
				&			&size			&				&			&size\\
\hline
\multicolumn{3}{c||}
{(a) nDCG}&
\multicolumn{3}{c}
{(b) Q}\\
\hline
PRI4-RND3 &0.0000000	&4.57	&RND3-PRI3 &0.00044	&2.22\\
PRI2-RND3 &0.0000001	&3.35	&PRI4-PRI3 &0.0021	&1.99\\
PRI1-RND3 &0.0000001	&3.34	&RND4-PRI3 &0.0058	&1.84\\
PRI3-RND3 &0.0000003	&3.14	&PRI2-PRI3 &0.025	&1.61\\
PRI4-RND4 &0.0000045	&2.82	&PRI1-PRI3 &0.047	&1.49\\
RND1-RND3 &0.000016		&2.65	&			&				&\\
PRI4-RND2 &0.00016		&2.35	&			&				&\\
RND2-RND3 &0.00041		&2.22	&			&				&\\
PRI4-RND1 &0.0035		&1.92	&			&				&\\
RND4-RND3 &0.010		&1.76	&			&				&\\
PRI2-RND4 &0.027		&1.60	&			&				&\\
PRI1-RND4 &0.029		&1.58	&			&				&\\
\hline
\multicolumn{3}{c||}
{(c) nERR}&
\multicolumn{3}{c}
{(d) iRBU}\\
\hline
			&			&		&PRI3-PRI2 &0.00023	&2.30\\
			&			&		&PRI3-RND1 &0.0032	&1.93\\
			&			&		&RND3-PRI2 &0.0092	&1.77\\
			&			&		&RND2-PRI2 &0.011	&1.74\\
\hline
\end{tabular}

\end{small}
\end{table}

\clearpage

\section{Conclusions}\label{s:conclusions}

The present study addressed a few questions that remained open for the past two decades or so
regarding 
two document ordering strategies for relevance assessors:
PRI (practiced at NTCIR) and
RND (recommended elsewhere, e.g., at TREC).
Our experiments, which involved eight independent relevance assessments 
for 32,375 topic-document pairs (i.e., a total of 259,000 labels),
provide some answers to them.
Our conclusions are as follows.
\begin{description}
\item[RQ1] 
\emph{Which strategy enables more efficient relevance assessments?}
There is no substantial difference between RND and PRI
in terms of time spent for judging each document,
although PRI may enable
faster identification of the first highly relevant document in the pool.
\item[RQ2] 
\emph{Which strategy enables higher inter-assessor agreements?}
The difference between
the inter-assessor agreement under the RND condition
and that under the PRI condition is probably of no practical significance.
\item[RQ3] 
\emph{Which strategy enables more stable system rankings across different versions of qrels files?}
While PRI-based qrels files tend to generate system ranking that are slightly more similar to each other
than RND-based qrels files do,
this difference is probably of no practical significance.
On the other hand, 
a PRI-based system ranking and a RND-based system ranking
can be quite different relative to the above within-PRI and within-RND ranking comparisons.
The PRI strategy tends to make the assessor favour ``popular'' documents,
i.e., those returned at high ranks by many systems.
\item[RQ4] 
\emph{Which strategy is more robust to systems that did not contribute to the pools?}
PRI-based qrels files tend to be slightly more robust to new systems than RND-based ones.
This is probably because the PRI strategy tends to help us identify ``popular'' relevant documents.
The ``popular'' relevant documents affect the evaluation of many systems,
including systems that did not contribute to the pools.
\end{description}
Also, additional experiments suggest that
the findings for \textbf{RQ3} and \textbf{RQ4}
may generalise to some extent to pool depths smaller than 15.

While we refrain from strongly recommending one document strategy over the other,
IR test collections builders and users should at least be aware that
(a)~document presentation order for the relevance assessors 
do affect which documents are judged (highly) relevant as well as 
the system ranking to some extent; and 
(b)~while the PRI strategy is probably biased towards 
documents that are returned by many systems at high ranks,
this may also provide a little more robustness to the handling of new systems (relative to RND),
at least if the new systems are quite similar to those that contributed documents 
to the pools.
That is, our results suggests that PRI-based test collections
may be slightly more reusable than RND-based ones.

One substantial limitation of the present study is that 
our data set is a large collection of  \emph{bronze assessor} labels:
the relevance assessors were students, not topic originators.
Hence, we have recently examined the effect of document ordering strategies on 
\emph{gold assessor} labels, i.e., those 
obtained from the topic originators~\cite{bailey08} (also called \emph{query owners}~\cite{chouldechova13}),
through construction of the NTCIR-16 WWW-4 test collection.
The results will be reported elsewhere.

\clearpage

\bibliographystyle{ACM-Reference-Format}
\bibliography{tois2022PRIRND-fullyrevised-arxiv}


\begin{thebibliography}{47}

%%% the Web address.

\ifx \showCODEN    \undefined \def \showCODEN     #1{\unskip}     \fi
\ifx \showDOI      \undefined \def \showDOI       #1{#1}\fi
\ifx \showISBNx    \undefined \def \showISBNx     #1{\unskip}     \fi
\ifx \showISBNxiii \undefined \def \showISBNxiii  #1{\unskip}     \fi
\ifx \showISSN     \undefined \def \showISSN      #1{\unskip}     \fi
\ifx \showLCCN     \undefined \def \showLCCN      #1{\unskip}     \fi
\ifx \shownote     \undefined \def \shownote      #1{#1}          \fi
\ifx \showarticletitle \undefined \def \showarticletitle #1{#1}   \fi
\ifx \showURL      \undefined \def \showURL       {\relax}        \fi
% invisible to TeX
\providecommand\bibfield[2]{#2}
\providecommand\bibinfo[2]{#2}
\providecommand\natexlab[1]{#1}
\providecommand\showeprint[2][]{arXiv:#2}

\bibitem[\protect\citeauthoryear{Allan, Carterette, Aslam, Pavlu, Dachev, and
  Kanoulas}{Allan et~al\mbox{.}}{2008}]%
        {allan08}
\bibfield{author}{\bibinfo{person}{James Allan}, \bibinfo{person}{Ben
  Carterette}, \bibinfo{person}{Javed~A. Aslam}, \bibinfo{person}{Virgil
  Pavlu}, \bibinfo{person}{Blagovest Dachev}, {and} \bibinfo{person}{Evangelos
  Kanoulas}.} \bibinfo{year}{2008}\natexlab{}.
\newblock \showarticletitle{Million Query Track 2007 Overview}.
\newblock


\bibitem[\protect\citeauthoryear{Allan, Harman, Kanoulas, Li, {Van Gysel}, and
  Voorhees}{Allan et~al\mbox{.}}{2018}]%
        {allan18}
\bibfield{author}{\bibinfo{person}{James Allan}, \bibinfo{person}{Donna
  Harman}, \bibinfo{person}{Evangelos Kanoulas}, \bibinfo{person}{Dan Li},
  \bibinfo{person}{Christophe {Van Gysel}}, {and} \bibinfo{person}{Ellen
  Voorhees}.} \bibinfo{year}{2018}\natexlab{}.
\newblock \showarticletitle{{TREC} Common Core Track Overview}. In
  \bibinfo{booktitle}{\emph{Proceedings of TREC 2017}}.
\newblock


\bibitem[\protect\citeauthoryear{Bailey, Craswell, Soboroff, Thomas, {de
  Vries}, and Yilmaz}{Bailey et~al\mbox{.}}{2008}]%
        {bailey08}
\bibfield{author}{\bibinfo{person}{Peter Bailey}, \bibinfo{person}{Nick
  Craswell}, \bibinfo{person}{Ian Soboroff}, \bibinfo{person}{Paul Thomas},
  \bibinfo{person}{Arjen~P. {de Vries}}, {and} \bibinfo{person}{Emine Yilmaz}.}
  \bibinfo{year}{2008}\natexlab{}.
\newblock \showarticletitle{Relevance Assessment: Are Judges Exchangeable and
  Does It Matter?}. In \bibinfo{booktitle}{\emph{Proceedings of ACM SIGIR
  2008}}. \bibinfo{pages}{667--674}.
\newblock


\bibitem[\protect\citeauthoryear{Carterette, Pavlu, Fang, and
  Kanoulas}{Carterette et~al\mbox{.}}{2010}]%
        {carterette10trec}
\bibfield{author}{\bibinfo{person}{Ben Carterette}, \bibinfo{person}{Virgil
  Pavlu}, \bibinfo{person}{Hui Fang}, {and} \bibinfo{person}{Evangelos
  Kanoulas}.} \bibinfo{year}{2010}\natexlab{}.
\newblock \showarticletitle{Million Query Track 2009 Overview}. In
  \bibinfo{booktitle}{\emph{Proceedings of TREC 2009}}.
\newblock


\bibitem[\protect\citeauthoryear{Chouldechova and Mease}{Chouldechova and
  Mease}{2013}]%
        {chouldechova13}
\bibfield{author}{\bibinfo{person}{Alexandra Chouldechova} {and}
  \bibinfo{person}{David Mease}.} \bibinfo{year}{2013}\natexlab{}.
\newblock \showarticletitle{Differences in Search Engine Evaluations Between
  Query Owners and Non-Owners}. In \bibinfo{booktitle}{\emph{Proceedings of ACM
  WSDM 2013}}. \bibinfo{pages}{103--112}.
\newblock


\bibitem[\protect\citeauthoryear{Clarke, Craswell, and Soboroff}{Clarke
  et~al\mbox{.}}{2010}]%
        {clarke10}
\bibfield{author}{\bibinfo{person}{Charles~L.A. Clarke}, \bibinfo{person}{Nick
  Craswell}, {and} \bibinfo{person}{Ian Soboroff}.}
  \bibinfo{year}{2010}\natexlab{}.
\newblock \showarticletitle{Overview of the {TREC} 2009 Web Track}. In
  \bibinfo{booktitle}{\emph{Proceedings of TREC 2009}}.
\newblock


\bibitem[\protect\citeauthoryear{Cleverdon and Keen}{Cleverdon and
  Keen}{1966}]%
        {cleverdon66b}
\bibfield{author}{\bibinfo{person}{Cyril Cleverdon} {and}
  \bibinfo{person}{Michael Keen}.} \bibinfo{year}{1966}\natexlab{}.
\newblock \bibinfo{booktitle}{\emph{Factors Determining the Performance of
  Indexing Systems; Volume 2}}.
\newblock \bibinfo{type}{{T}echnical {R}eport}. \bibinfo{institution}{College
  of Aeronautics, Cranfield, UK}.
\newblock


\bibitem[\protect\citeauthoryear{Cleverdon, Mills, and Keen}{Cleverdon
  et~al\mbox{.}}{1966}]%
        {cleverdon66a}
\bibfield{author}{\bibinfo{person}{Cyril Cleverdon}, \bibinfo{person}{Jack
  Mills}, {and} \bibinfo{person}{Michael Keen}.}
  \bibinfo{year}{1966}\natexlab{}.
\newblock \bibinfo{booktitle}{\emph{Factors Determining the Performance of
  Indexing Systems; Volume 1: Design}}.
\newblock \bibinfo{type}{{T}echnical {R}eport}. \bibinfo{institution}{College
  of Aeronautics, Cranfield, UK}.
\newblock


\bibitem[\protect\citeauthoryear{Cormack, Palmer, and Clarke}{Cormack
  et~al\mbox{.}}{1998}]%
        {cormack98}
\bibfield{author}{\bibinfo{person}{Gordon~V. Cormack},
  \bibinfo{person}{Christopher~R. Palmer}, {and} \bibinfo{person}{Charles~L.A.
  Clarke}.} \bibinfo{year}{1998}\natexlab{}.
\newblock \showarticletitle{Efficient Construction of Large Test Collections}.
  In \bibinfo{booktitle}{\emph{Proceedings of ACM SIGIR '98}}.
  \bibinfo{pages}{282--289}.
\newblock


\bibitem[\protect\citeauthoryear{Damessie, Culpepper, Kim, and
  Scholer}{Damessie et~al\mbox{.}}{2018}]%
        {damessie18}
\bibfield{author}{\bibinfo{person}{Tadele~T. Damessie},
  \bibinfo{person}{J.~Shane Culpepper}, \bibinfo{person}{Jaewon Kim}, {and}
  \bibinfo{person}{Falk Scholer}.} \bibinfo{year}{2018}\natexlab{}.
\newblock \showarticletitle{Presentation Ordering Effects on Assessor
  Agreement}. In \bibinfo{booktitle}{\emph{Proceedings of ACM CIKM 2018}}.
  \bibinfo{pages}{723--732}.
\newblock


\bibitem[\protect\citeauthoryear{Damessie, Scholer, J\"{a}rvelin, and
  Culpepper}{Damessie et~al\mbox{.}}{2016}]%
        {damessie16}
\bibfield{author}{\bibinfo{person}{Tadele~T. Damessie}, \bibinfo{person}{Falk
  Scholer}, \bibinfo{person}{Kalervo J\"{a}rvelin}, {and}
  \bibinfo{person}{J.~Shane Culpepper}.} \bibinfo{year}{2016}\natexlab{}.
\newblock \showarticletitle{The Effect of Document Order and Topic Difficulty
  on Assessor Agreement}. In \bibinfo{booktitle}{\emph{Proceedings of ACM ICTIR
  2016}}. \bibinfo{pages}{73--76}.
\newblock


\bibitem[\protect\citeauthoryear{Eisenberg and Barry}{Eisenberg and
  Barry}{1988}]%
        {eisenberg88}
\bibfield{author}{\bibinfo{person}{Michael Eisenberg} {and}
  \bibinfo{person}{Carol Barry}.} \bibinfo{year}{1988}\natexlab{}.
\newblock \showarticletitle{Order Effects: A Study of the Possible Influence of
  Presentation Order on User Judgments of Document Relevance}.
\newblock \bibinfo{journal}{\emph{Journal of the American Society for
  Information Science}} \bibinfo{volume}{39}, \bibinfo{number}{5}
  (\bibinfo{year}{1988}), \bibinfo{pages}{293--300}.
\newblock


\bibitem[\protect\citeauthoryear{Ferro and Peters}{Ferro and Peters}{2019}]%
        {ferro19book}
\bibfield{editor}{\bibinfo{person}{Nicola Ferro} {and} \bibinfo{person}{Carol
  Peters}} (Eds.). \bibinfo{year}{2019}\natexlab{}.
\newblock \bibinfo{booktitle}{\emph{Information Retrieval in a Changing World:
  Lessons Learned from 20 Years of {CLEF}}}.
\newblock \bibinfo{publisher}{Springer}.
\newblock


\bibitem[\protect\citeauthoryear{Harman}{Harman}{2005}]%
        {harman05}
\bibfield{author}{\bibinfo{person}{Donna~K. Harman}.}
  \bibinfo{year}{2005}\natexlab{}.
\newblock \showarticletitle{The {TREC} Test Collections}.
\newblock In \bibinfo{booktitle}{\emph{TREC: Experiment and Evaluation in
  Information Retrieval}}, \bibfield{editor}{\bibinfo{person}{Ellen~M.
  Voorhees} {and} \bibinfo{person}{Donna~K. Harman}} (Eds.).
  \bibinfo{publisher}{The MIT Press}, Chapter~2.
\newblock


\bibitem[\protect\citeauthoryear{Huang and Wang}{Huang and Wang}{2004}]%
        {huang04}
\bibfield{author}{\bibinfo{person}{Mu-Hsuan Huang} {and}
  \bibinfo{person}{Hui-Yu Wang}.} \bibinfo{year}{2004}\natexlab{}.
\newblock \showarticletitle{The Influence of Document Presentation Order and
  Number of Documents Judged on Users' Judgments of Relevance}.
\newblock \bibinfo{journal}{\emph{Journal of the American Society for
  Information Science}} \bibinfo{volume}{55}, \bibinfo{number}{11}
  (\bibinfo{year}{2004}), \bibinfo{pages}{970--979}.
\newblock


\bibitem[\protect\citeauthoryear{Kando}{Kando}{2004}]%
        {kando04}
\bibfield{author}{\bibinfo{person}{Noriko Kando}.}
  \bibinfo{year}{2004}\natexlab{}.
\newblock \showarticletitle{Evaluation of Information Access Technologies at
  the {NTCIR} Workshop}. In \bibinfo{booktitle}{\emph{Proceedings of CLEF 2003
  (LNCS 3237)}}. \bibinfo{pages}{29--43}.
\newblock


\bibitem[\protect\citeauthoryear{Krippendorff}{Krippendorff}{2018}]%
        {krippendorff18}
\bibfield{author}{\bibinfo{person}{Klaus Krippendorff}.}
  \bibinfo{year}{2018}\natexlab{}.
\newblock \bibinfo{booktitle}{\emph{Content Analysis: An Introduction to Its
  Methodology (Fourth Edition)}}.
\newblock \bibinfo{publisher}{SAGE Publications}.
\newblock


\bibitem[\protect\citeauthoryear{Lipani, Losada, Zuccon, and Lupu}{Lipani
  et~al\mbox{.}}{2021}]%
        {lipani21}
\bibfield{author}{\bibinfo{person}{Aldo Lipani}, \bibinfo{person}{David~E.
  Losada}, \bibinfo{person}{Guido Zuccon}, {and} \bibinfo{person}{Mihai Lupu}.}
  \bibinfo{year}{2021}\natexlab{}.
\newblock \showarticletitle{Fixed-Cost Pooling Strategies}.
\newblock \bibinfo{journal}{\emph{IEEE Transactions on Knowledge and Data
  Engineering}} \bibinfo{volume}{33}, \bibinfo{number}{4}
  (\bibinfo{year}{2021}), \bibinfo{pages}{1503--1522}.
\newblock


\bibitem[\protect\citeauthoryear{Losada, Parapar, and \'{A}lvaro
  Barreiro}{Losada et~al\mbox{.}}{2017}]%
        {losada17}
\bibfield{author}{\bibinfo{person}{David~E. Losada}, \bibinfo{person}{Javier
  Parapar}, {and} \bibinfo{person}{\'{A}lvaro Barreiro}.}
  \bibinfo{year}{2017}\natexlab{}.
\newblock \showarticletitle{Multi-armed Bandits for Ordering Judgements in
  Pooling-based Evaluation}.
\newblock \bibinfo{journal}{\emph{Information Processing and Management}}
  \bibinfo{volume}{53}, \bibinfo{number}{3} (\bibinfo{year}{2017}),
  \bibinfo{pages}{1005--1025}.
\newblock


\bibitem[\protect\citeauthoryear{Losada, Parapar, and \'{A}lvaro
  Barreiro}{Losada et~al\mbox{.}}{2018}]%
        {losada18}
\bibfield{author}{\bibinfo{person}{David~E. Losada}, \bibinfo{person}{Javier
  Parapar}, {and} \bibinfo{person}{\'{A}lvaro Barreiro}.}
  \bibinfo{year}{2018}\natexlab{}.
\newblock \showarticletitle{When to Stop Making Relevance Judgments? A Study of
  Stopping Methods for Building Information Retrieval Test Collections}.
\newblock \bibinfo{journal}{\emph{Journal of the Association for Information
  Science and Technology}} (\bibinfo{year}{2018}).
\newblock


\bibitem[\protect\citeauthoryear{Luo, Sakai, Liu, Dou, Xiong, and Xu}{Luo
  et~al\mbox{.}}{2017}]%
        {luo17}
\bibfield{author}{\bibinfo{person}{Cheng Luo}, \bibinfo{person}{Tetsuya Sakai},
  \bibinfo{person}{Yiqun Liu}, \bibinfo{person}{Zhicheng Dou},
  \bibinfo{person}{Chenyan Xiong}, {and} \bibinfo{person}{Jingfang Xu}.}
  \bibinfo{year}{2017}\natexlab{}.
\newblock \showarticletitle{Overview of the {NTCIR}-13 We Want Web Task}. In
  \bibinfo{booktitle}{\emph{Proceedings of NTCIR-13}}.
  \bibinfo{pages}{394--401}.
\newblock


\bibitem[\protect\citeauthoryear{Mao, Sakai, Luo, Xiao, Liu, and Dou}{Mao
  et~al\mbox{.}}{2019}]%
        {mao19}
\bibfield{author}{\bibinfo{person}{Jiaxin Mao}, \bibinfo{person}{Tetsuya
  Sakai}, \bibinfo{person}{Cheng Luo}, \bibinfo{person}{Peng Xiao},
  \bibinfo{person}{Yiqun Liu}, {and} \bibinfo{person}{Zhicheng Dou}.}
  \bibinfo{year}{2019}\natexlab{}.
\newblock \showarticletitle{Overview of the {NTCIR}-14 We Want Web Task}. In
  \bibinfo{booktitle}{\emph{Proceedings of NTCIR-14}}.
  \bibinfo{pages}{455--467}.
\newblock


\bibitem[\protect\citeauthoryear{Sakai}{Sakai}{2008}]%
        {sakai08cikm}
\bibfield{author}{\bibinfo{person}{Tetsuya Sakai}.}
  \bibinfo{year}{2008}\natexlab{}.
\newblock \showarticletitle{Comparing Metrics across TREC and NTCIR: The
  Robustness to System Bias}. In \bibinfo{booktitle}{\emph{Proceedings of ACM
  CIKM 2008}}. \bibinfo{pages}{581--590}.
\newblock


\bibitem[\protect\citeauthoryear{Sakai}{Sakai}{2014}]%
        {sakai14promise}
\bibfield{author}{\bibinfo{person}{Tetsuya Sakai}.}
  \bibinfo{year}{2014}\natexlab{}.
\newblock \showarticletitle{Metrics, Statistics, Tests}. In
  \bibinfo{booktitle}{\emph{PROMISE Winter School 2013: Bridging between
  Information Retrieval and Databases (LNCS 8173)}}. \bibinfo{pages}{116--163}.
\newblock


\bibitem[\protect\citeauthoryear{Sakai}{Sakai}{2016a}]%
        {sakai16sigir}
\bibfield{author}{\bibinfo{person}{Tetsuya Sakai}.}
  \bibinfo{year}{2016}\natexlab{a}.
\newblock \showarticletitle{Statistical Significance, Power, and Sample Sizes:
  A Systematic Review of {SIGIR} and {TOIS}, 2006-2015}. In
  \bibinfo{booktitle}{\emph{Proceedings of ACM SIGIR 2016}}.
  \bibinfo{pages}{5--14}.
\newblock


\bibitem[\protect\citeauthoryear{Sakai}{Sakai}{2016b}]%
        {sakai16irj}
\bibfield{author}{\bibinfo{person}{Tetsuya Sakai}.}
  \bibinfo{year}{2016}\natexlab{b}.
\newblock \showarticletitle{Topic Set Size Design}.
\newblock \bibinfo{journal}{\emph{Information Retrieval Journal}}
  \bibinfo{volume}{19}, \bibinfo{number}{3} (\bibinfo{year}{2016}),
  \bibinfo{pages}{256--283}.
\newblock


\bibitem[\protect\citeauthoryear{Sakai}{Sakai}{2018}]%
        {sakai18book}
\bibfield{author}{\bibinfo{person}{Tetsuya Sakai}.}
  \bibinfo{year}{2018}\natexlab{}.
\newblock \showarticletitle{Laboratory Experiments in Information Retrieval:
  Sample Sizes, Effect Sizes, and Statistical Power}.
  \bibinfo{publisher}{Springer}.
\newblock


\bibitem[\protect\citeauthoryear{Sakai}{Sakai}{2019}]%
        {sakai19clefbook}
\bibfield{author}{\bibinfo{person}{Tetsuya Sakai}.}
  \bibinfo{year}{2019}\natexlab{}.
\newblock \showarticletitle{How to Run an Evaluation Task}.
\newblock In \bibinfo{booktitle}{\emph{Information Retrieval Evaluation in a
  Changing World}}, \bibfield{editor}{\bibinfo{person}{Nicola Ferro} {and}
  \bibinfo{person}{Carol Peters}} (Eds.). \bibinfo{publisher}{Springer}.
\newblock


\bibitem[\protect\citeauthoryear{Sakai}{Sakai}{2021}]%
        {sakai21ecir}
\bibfield{author}{\bibinfo{person}{Tetsuya Sakai}.}
  \bibinfo{year}{2021}\natexlab{}.
\newblock \showarticletitle{On the Instability of Diminishing Return IR
  Measures}. In \bibinfo{booktitle}{\emph{Proceedings of ECIR 2021 Part I (LNCS
  12656)}}. \bibinfo{pages}{572--586}.
\newblock


\bibitem[\protect\citeauthoryear{Sakai, Dou, Song, and Kando}{Sakai
  et~al\mbox{.}}{2012}]%
        {sakai12airs}
\bibfield{author}{\bibinfo{person}{Tetsuya Sakai}, \bibinfo{person}{Zhicheng
  Dou}, \bibinfo{person}{Ruihua Song}, {and} \bibinfo{person}{Noriko Kando}.}
  \bibinfo{year}{2012}\natexlab{}.
\newblock \showarticletitle{The Reusability of a Diversified Search Test
  Collection}. In \bibinfo{booktitle}{\emph{Proceedings of AIRS 2012 (LNCS
  7675)}}. \bibinfo{pages}{26--38}.
\newblock


\bibitem[\protect\citeauthoryear{Sakai, Kando, Lin, Mitamura, Shima, Ji, Chen,
  and Nyberg}{Sakai et~al\mbox{.}}{2008}]%
        {sakai08ir4qa}
\bibfield{author}{\bibinfo{person}{Tetsuya Sakai}, \bibinfo{person}{Noriko
  Kando}, \bibinfo{person}{Chuan-Jie Lin}, \bibinfo{person}{Teruko Mitamura},
  \bibinfo{person}{Hideki Shima}, \bibinfo{person}{Donghong Ji},
  \bibinfo{person}{Kuang-Hua Chen}, {and} \bibinfo{person}{Eric Nyberg}.}
  \bibinfo{year}{2008}\natexlab{}.
\newblock \showarticletitle{Overview of the {NTCIR}-7 {ACLIA} {IR4QA} Task}. In
  \bibinfo{booktitle}{\emph{Proceedings of NTCIR-7}}. \bibinfo{pages}{77--114}.
\newblock


\bibitem[\protect\citeauthoryear{Sakai and Lin}{Sakai and Lin}{2010}]%
        {sakai10evia}
\bibfield{author}{\bibinfo{person}{Tetsuya Sakai} {and}
  \bibinfo{person}{Chin-Yew Lin}.} \bibinfo{year}{2010}\natexlab{}.
\newblock \showarticletitle{Ranking Retrieval Systems without Relevance
  Assessments - Revisited}. In \bibinfo{booktitle}{\emph{Proceedings of EVIA
  2010}}. \bibinfo{pages}{25--33}.
\newblock


\bibitem[\protect\citeauthoryear{Sakai, Oard, and Kando}{Sakai
  et~al\mbox{.}}{2020a}]%
        {sakai20book}
\bibfield{editor}{\bibinfo{person}{Tetsuya Sakai}, \bibinfo{person}{Douglas~W.
  Oard}, {and} \bibinfo{person}{Noriko Kando}} (Eds.).
  \bibinfo{year}{2020}\natexlab{a}.
\newblock \bibinfo{booktitle}{\emph{Evaluating Information Retrieval and Access
  Tasks: {NTCIR}'s Legacy of Research Impact}}.
\newblock \bibinfo{publisher}{Springer}.
\newblock


\bibitem[\protect\citeauthoryear{Sakai, Tao, Maistro, Chu, Li, Chen, Ferro,
  Wang, Soboroff, and Liu}{Sakai et~al\mbox{.}}{2022b}]%
        {sakai22www234corrected}
\bibfield{author}{\bibinfo{person}{Tetsuya Sakai}, \bibinfo{person}{Sijie Tao},
  \bibinfo{person}{Maria Maistro}, \bibinfo{person}{Zhumin Chu},
  \bibinfo{person}{Yujing Li}, \bibinfo{person}{Nuo Chen},
  \bibinfo{person}{Nicola Ferro}, \bibinfo{person}{Junjie Wang},
  \bibinfo{person}{Ian Soboroff}, {and} \bibinfo{person}{Yiqun Liu}.}
  \bibinfo{year}{2022}\natexlab{b}.
\newblock \showarticletitle{Corrected Evaluation Results of the {NTCIR}
  {WWW}-2, {WWW}-3, and {WWW}-4 English Subtasks}.
\newblock  (\bibinfo{year}{2022}).
\newblock
\urldef\tempurl%
\url{http://arxiv.org/abs/2210.10266}
\showURL{%
\tempurl}


\bibitem[\protect\citeauthoryear{Sakai, Tao, and Zeng}{Sakai
  et~al\mbox{.}}{2021}]%
        {sakai21www3e8}
\bibfield{author}{\bibinfo{person}{Tetsuya Sakai}, \bibinfo{person}{Sijie Tao},
  {and} \bibinfo{person}{Zhaohao Zeng}.} \bibinfo{year}{2021}\natexlab{}.
\newblock \showarticletitle{WWW3E8: 259,000 Relevance Labels for Studying the
  Effect of Document Presentation Order for Relevance Assessors}. In
  \bibinfo{booktitle}{\emph{Proceedings of ACM SIGIR 2021}}. \bibinfo{pages}{to
  appear}.
\newblock


\bibitem[\protect\citeauthoryear{Sakai, Tao, and Zeng}{Sakai
  et~al\mbox{.}}{2022a}]%
        {sakai22tois}
\bibfield{author}{\bibinfo{person}{Tetsuya Sakai}, \bibinfo{person}{Sijie Tao},
  {and} \bibinfo{person}{Zhaohao Zeng}.} \bibinfo{year}{2022}\natexlab{a}.
\newblock \showarticletitle{Relevance Assessments for Web Search Evaluation:
  Should We Randomise or Prioritise the Pooled Documents?}
\newblock \bibinfo{journal}{\emph{ACM TOIS}} \bibinfo{volume}{40},
  \bibinfo{number}{4}, Article \bibinfo{articleno}{76} (\bibinfo{year}{2022}).
\newblock


\bibitem[\protect\citeauthoryear{Sakai, Tao, Zeng, Zheng, Mao, Chu, Liu, Dou,
  Ferro, Maistro, and Soboroff}{Sakai et~al\mbox{.}}{2020b}]%
        {sakai20www3}
\bibfield{author}{\bibinfo{person}{Tetsuya Sakai}, \bibinfo{person}{Sijie Tao},
  \bibinfo{person}{Zhaohao Zeng}, \bibinfo{person}{Yukun Zheng},
  \bibinfo{person}{Jiaxin Mao}, \bibinfo{person}{Zhumin Chu},
  \bibinfo{person}{Yiqun Liu}, \bibinfo{person}{Zhicheng Dou},
  \bibinfo{person}{Nicola Ferro}, \bibinfo{person}{Maria Maistro}, {and}
  \bibinfo{person}{Ian Soboroff}.} \bibinfo{year}{2020}\natexlab{b}.
\newblock \showarticletitle{Overview of the {NTCIR-15 We Want Web with CENTRE
  (WWW-3)} Task}. In \bibinfo{booktitle}{\emph{Proceedings of NTCIR-15}}.
  \bibinfo{pages}{219--234}.
\newblock


\bibitem[\protect\citeauthoryear{Sakai and Xiao}{Sakai and Xiao}{2019}]%
        {sakai19airs}
\bibfield{author}{\bibinfo{person}{Tetsuya Sakai} {and} \bibinfo{person}{Peng
  Xiao}.} \bibinfo{year}{2019}\natexlab{}.
\newblock \showarticletitle{Randomised vs. Prioritised Pools for Relevance
  Assessments: Sample Size Considerations}. In
  \bibinfo{booktitle}{\emph{Proceedings of AIRS 2019 (LNCS 12004)}}.
  \bibinfo{pages}{94--105}.
\newblock


\bibitem[\protect\citeauthoryear{Sakai and Zeng}{Sakai and Zeng}{2019}]%
        {sakai19sigir}
\bibfield{author}{\bibinfo{person}{Tetsuya Sakai} {and}
  \bibinfo{person}{Zhaohao Zeng}.} \bibinfo{year}{2019}\natexlab{}.
\newblock \showarticletitle{Which Diversity Evaluation Measures are ``Good''?}.
  In \bibinfo{booktitle}{\emph{Proceedings of ACM SIGIR 2019}}.
  \bibinfo{pages}{595--604}.
\newblock


\bibitem[\protect\citeauthoryear{Sakai and Zeng}{Sakai and Zeng}{2020}]%
        {sakai20tois}
\bibfield{author}{\bibinfo{person}{Tetsuya Sakai} {and}
  \bibinfo{person}{Zhaohao Zeng}.} \bibinfo{year}{2020}\natexlab{}.
\newblock \showarticletitle{Retrieval Evaluation Measures that Agree with
  Users’ SERP Preferences: Traditional, Preference-based, and Diversity
  Measures}.
\newblock \bibinfo{journal}{\emph{ACM TOIS}} \bibinfo{volume}{39},
  \bibinfo{number}{2}, Article \bibinfo{articleno}{14} (\bibinfo{year}{2020}).
\newblock


\bibitem[\protect\citeauthoryear{Scholer, Kelly, Wu, Lee, and Webber}{Scholer
  et~al\mbox{.}}{2013}]%
        {scholer13}
\bibfield{author}{\bibinfo{person}{Falk Scholer}, \bibinfo{person}{Diane
  Kelly}, \bibinfo{person}{Wan-Ching Wu}, \bibinfo{person}{Hanseul~S. Lee},
  {and} \bibinfo{person}{William Webber}.} \bibinfo{year}{2013}\natexlab{}.
\newblock \showarticletitle{The Effect of Threshold Priming and Need for
  Cognition on Relevance Calibration and Assessment}. In
  \bibinfo{booktitle}{\emph{Proceedings of ACM SIGIR 2013}}.
  \bibinfo{pages}{623--632}.
\newblock


\bibitem[\protect\citeauthoryear{{Sparck Jones} and Bates}{{Sparck Jones} and
  Bates}{1977}]%
        {sparckjones77}
\bibfield{author}{\bibinfo{person}{K. {Sparck Jones}} {and}
  \bibinfo{person}{R.~G. Bates}.} \bibinfo{year}{1977}\natexlab{}.
\newblock \bibinfo{booktitle}{\emph{Report on a Design Study for the 'Ideal'
  information Retrieval Test Collection}}.
\newblock \bibinfo{type}{{T}echnical {R}eport}. \bibinfo{institution}{Computer
  Laboratory, University of Cambridge, British Library Research and Development
  Report No.5481}.
\newblock


\bibitem[\protect\citeauthoryear{{Sparck Jones} and {van Rijsbergen}}{{Sparck
  Jones} and {van Rijsbergen}}{1975}]%
        {sparckjones75}
\bibfield{author}{\bibinfo{person}{K. {Sparck Jones}} {and}
  \bibinfo{person}{C.~J. {van Rijsbergen}}.} \bibinfo{year}{1975}\natexlab{}.
\newblock \bibinfo{booktitle}{\emph{Report on the Need for and Provision of an
  'Ideal' Information Retrieval Test Collection}}.
\newblock \bibinfo{type}{{T}echnical {R}eport}. \bibinfo{institution}{Computer
  Laboratory, University of Cambridge, British Library Research and Development
  Report No.5266}.
\newblock


\bibitem[\protect\citeauthoryear{Turpin, Scholer, Mizzaro, and
  Maddalena}{Turpin et~al\mbox{.}}{2015}]%
        {turpin15}
\bibfield{author}{\bibinfo{person}{Andrew Turpin}, \bibinfo{person}{Falk
  Scholer}, \bibinfo{person}{Stefano Mizzaro}, {and} \bibinfo{person}{Eddy
  Maddalena}.} \bibinfo{year}{2015}\natexlab{}.
\newblock \showarticletitle{The Benefits of Magnitude Estimation Relevance
  Assessments for Information Retrieval Evaluation}. In
  \bibinfo{booktitle}{\emph{Proceedings of ACM SIGIR 2015}}.
  \bibinfo{pages}{565--574}.
\newblock


\bibitem[\protect\citeauthoryear{Voorhees}{Voorhees}{2000}]%
        {voorhees00}
\bibfield{author}{\bibinfo{person}{Ellen~M. Voorhees}.}
  \bibinfo{year}{2000}\natexlab{}.
\newblock \showarticletitle{Variations in Relevance Judgments and the
  Measurement of Retrieval Effectiveness}.
\newblock \bibinfo{journal}{\emph{Information Processing and Management}}
  \bibinfo{volume}{36} (\bibinfo{year}{2000}), \bibinfo{pages}{697--716}.
\newblock


\bibitem[\protect\citeauthoryear{Voorhees}{Voorhees}{2002}]%
        {voorhees02}
\bibfield{author}{\bibinfo{person}{Ellen~M. Voorhees}.}
  \bibinfo{year}{2002}\natexlab{}.
\newblock \showarticletitle{The Philosophy of Information Retrieval
  Evaluation}. In \bibinfo{booktitle}{\emph{Proceedings of CLEF 2001 (LNCS
  2406)}}. \bibinfo{pages}{355--370}.
\newblock


\bibitem[\protect\citeauthoryear{Zobel}{Zobel}{1998}]%
        {zobel98}
\bibfield{author}{\bibinfo{person}{Justin Zobel}.}
  \bibinfo{year}{1998}\natexlab{}.
\newblock \showarticletitle{How Reliable are the Results of Large-Scale
  Information Retrieval Experiments?}. In \bibinfo{booktitle}{\emph{Proceedings
  of ACM SIGIR '98}}. \bibinfo{pages}{307--314}.
\newblock


\end{thebibliography}

\end{document}